\documentclass[aoas,MSNbibl,nameyear,dvips]{arximspdf}
\usepackage{mathbh}
\usepackage{mathrsfs}
\usepackage{graphicx}

\doi{10.1214/14-AOAS782} 
\volume{8}
\issue{4}
\pubyear{2014}
\firstpage{2027}
\lastpage{2067}
\docsubty{FLA}

\makeatletter
\newcommand{\ler}{\mathrm{le}}
\newcommand{\ri}{\mathrm{ri}}
\newcommand{\rrvert}{\vert}
\newcommand{\rrVert}{\Vert}
\newcommand{\llvert}{\vert}
\newcommand{\llVert}{\Vert}
\renewcommand{\mid}{|}
\newcommand{\nicefrac}[2]{#1/{#2}}
\newcommand{\Var}{\operatorname{Var}}
\newcommand{\mathbbm}{\mathbh}
\newtheorem{theo}{Theorem}[section]
\newtheorem{coro}[theo]{Corollary}
\newtheorem{prop}[theo]{Proposition}
\newproclaim{defi}[theo]{Definition}
\newtheorem{lemm}[theo]{Lemma}
\newtheorem{cons}[theo]{Construction}
\makeatother

\begin{document}
\begin{frontmatter}

\title{A multiple filter test for the detection of rate changes in
renewal processes with varying variance}
\runtitle{A multiple filter test}

\begin{aug}
\author[A]{\fnms{Michael}~\snm{Messer}\thanksref{M1,T1}\ead[label=e1]{messer@math.uni-frankfurt.de}},
\author[B]{\fnms{Marietta}~\snm{Kirchner}\ead[label=e2]{kirchner@imbi.uni-heidelberg.de}\thanksref{M2}},
\author[C]{\fnms{Julia}~\snm{Schiemann}\ead[label=e3]{Julia.Schiemann@ed.ac.uk}\thanksref{M1,M3}},
\author[D]{\fnms{Jochen}~\snm{Roeper}\thanksref{M1,T1,T2}\ead[label=e4]{roeper@em.uni-frankfurt.de}},
\author[A]{\fnms{Ralph}~\snm{Neininger}\ead[label=e5]{neiningr@math.uni-frankfurt.de}\thanksref{M1}}
\and
\author[A]{\fnms{Gaby}~\snm{Schneider}\corref{}\thanksref{T1,T2}\ead[label=e6]{schneider@math.uni-frankfurt.de}\ead[label=u1,url]{http://www.math.uni-frankfurt.de/\textasciitilde ismi/schneider/}\thanksref{M1}}
\runauthor{M. Messer et al.}
\affiliation{Goethe University Frankfurt\thanksmark{M1},
Heidelberg University\thanksmark{M2}\\
and University of Edinburgh\thanksmark{M3}}
\address[A]{M.~Messer\\
R. Neininger\\
G. Schneider\\
Institute of Mathematics\\
Goethe University Frankfurt\\
Robert-Mayer-Str. 10\\
60325 Frankfurt\\
Germany\\
\printead{e1}\\
\phantom{E-mail: }\printead*{e5}\\
\phantom{E-mail: }\printead*{e6}}
\address[B]{M. Kirchner\\
Institute of Medical Biometry\\
\quad and Informatics\\
Heidelberg University\\
Im Neuenheimer Feld 305\\
69120 Heidelberg\\
Germany\\
\printead{e2}}
\address[C]{J. Schiemann\\
Institute of Neurophysiology\\
Neuroscience Center\\
Goethe University Frankfurt\\
Theodor-Stern-Kai 7\\
60590 Frankfurt\\
Germany\\
and\\
Centre for Integrative Physiology\\
University of Edinburgh\\
George Square\\
Edinburgh EH8 9XD\\
United Kingdom\\
\printead{e3}}
\address[D]{J. Roeper\\
Institute of Neurophysiology\\
Neuroscience Center\\
Goethe University Frankfurt\\
Theodor-Stern-Kai 7\\
60590 Frankfurt\\
Germany\\
\printead{e4}\hspace*{16pt}}
\end{aug}
\thankstext{T1}{Supported in part by the LOEWE ``Neuronale Koordination
Forschungsschwerpunkt Frankfurt.''}
\thankstext{T2}{Supported in part by the Priority Program 1665 of the DFG.}

\received{\smonth{4} \syear{2013}}
\revised{\smonth{11} \syear{2013}}

%
\begin{abstract}
Nonstationarity of the event rate is a persistent problem in modeling
time series of events, such as neuronal spike trains.
Motivated by a variety of patterns in neurophysiological spike train
recordings, we define a general class of renewal
processes. This class is used to test the null hypothesis of stationary
rate versus a wide alternative of renewal processes with finitely many
rate changes (change points). Our test extends ideas from the filtered
derivative approach by using multiple moving windows simultaneously. To
adjust the rejection threshold of the test, we use a Gaussian process,
which emerges as the limit of the filtered derivative process. We also
develop a multiple filter algorithm, which can be used when the null
hypothesis is rejected in order to estimate the number and location of
change points. We analyze the benefits of multiple filtering and its
increased detection probability as compared to a single window
approach. Application to spike trains recorded from dopamine midbrain
neurons in anesthetized mice illustrates the relevance of the proposed
techniques as preprocessing steps for methods that assume rate
stationarity. In over 70\% of all analyzed spike trains classified as
rate nonstationary, different change points were detected by different
window sizes.
\end{abstract}

%
\end{frontmatter}

\section{Introduction}
In neurophysiology, spike trains are often analyzed with statistical
models based on point processes, for example, renewal processes
\citet{Perkel67a,Johnson1996,Rieke1997,KassEtal2005,Nawrot2008}. A
large field of statistical neuroscience focuses on the coordination
between parallel point processes [\citet
{Perkel67b,BrownKM04,GruenRotter2010}]. In many models used for such
analyses, rate stationarity is a crucial assumption, and variations of
the underlying firing rate can affect the results of the applied
techniques [e.g., \citet{Brody1999,GruenEtAl2003}]. In order to
avoid such problems, several authors have suggested local techniques,
which involve the separate treatment of sections with approximately
stationary rate [see, e.g., \citet
{Gruen02b,Staude2008,Schneider08}] when spike trains show nonstationary
properties. Therefore, it is important to capture these nonstationary
properties, that is, to detect the violation of rate stationarity and
to locate the changes in the firing rate of neurons.

\begin{figure}[b]

\includegraphics{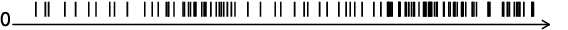}

\caption{A time series of events for which visual inspection suggests a
nonstationary rate. For
a general class of point processes, we present a statistical test and
an algorithm based on multiple windows in order to identify the number
and location of change points in the rate.}\label{motivation}
\end{figure}

In this paper we contribute to the change point analysis of point
processes. Motivated by the modeling of empirical data from
neurophysiology, we define a general class of renewal processes. In
this class, we test the null hypothesis of rate stationarity versus a
wide alternative of renewal processes with finitely many rate changes.
Our test extends ideas from the filtered derivative approach
[\citet{Steinebach1995,Bertrand2000}] by using multiple moving
windows simultaneously instead of just one moving window. To adjust the
rejection threshold of the test, we use a Gaussian process, which
emerges as the limit of the filtered derivative process. Additionally,
we develop a multiple filter algorithm, which can be used when the null
hypothesis is rejected in order to estimate the number and location of
change points. We analyze the benefits of our multiple filter algorithm
and study the increase in detection probability against single window
techniques.
This procedure can serve as a preprocessing step, splitting up the time
series into sections, in which the analyses of interest can be
performed separately. As an example, Figure~\ref{motivation}
illustrates a point process with nonstationary rate, in which we aim to
estimate the number and location of change points.

%

For identifying the number and positions of change points in time
series, many techniques are available in mathematical statistics. For
an overview see, for example, \citet{Basseville1993,Brodsky1993,Csorgo1998}. Typically, these techniques
are derived in the context of time series models with independent and
identically distributed (i.i.d.) random variables. The classical
parametric test uses a maximized likelihood quotient in order to
analyze the entire process, which leads to so-called pontograms in
point/renewal process theory [see \citeauthor{Csorgo1987} (\citeyear{Csorgo1987,Csorgo1998}),
\citet{kendall1980,Steinebach1993}]. The resulting test
statistics have extreme-value type limits [\citet{Huskova2001}].
As a second approach, moving window analyses in the context of renewal
processes have been studied by \citet{Steinebach1995}. These local
concepts successively investigate the life times of the point process
instead of referring to the entire process.

Motivated by applications, we present two extensions of existing
methods: first, the high variability of point processes observed
empirically requires a sufficiently general class of point process
models. Accordingly, we first introduce in Section~\ref{sectmodelass}
a new class of renewal processes with varying variance (RPVV), which
allow a certain variability in the variance of the life time
distributions. This generalization has the additional advantage that
rate changes can be investigated irrespective of variance changes and
that the latter could then be analyzed in a subsequent, separate
analysis which respects the identified rate changes. As a second
extension to existing methods, we take into account that rate changes
can occur on fast and slow time scales within the same time series. We
propose a multiple filter technique that applies multiple windows
simultaneously. This technique consists of a statistical multiple
filter test (MFT) for the null hypothesis of rate stationarity and a
multiple filter algorithm (MFA) for change point detection.

In Section~\ref{secmft} we first extend techniques introduced by
\citet{Steinebach1995} to our class of RPVVs. In particular, we
prove asymptotic results for a moving average approach called filtered
derivative, which is based on comparing the number of events in
adjacent windows. We then introduce a statistical test that is based on
a set of filtered derivative processes, each process corresponding to
one window size. The maximum over all processes serves as a test
statistic, indicating deviations from rate stationarity if this maximum
exceeds a threshold~$Q$. By scaling each process, we attempt to give
every window a similar impact on the maximum distribution.

For practical application, we provide in Section~\ref{seccpd} a
multiple filter algorithm for change point detection, in which the
results obtained by multiple window sizes are combined. For each
individual window, the algorithm successively searches for extreme
values of the filtered derivative, similar to the techniques proposed
by \citet{Bertrand2000,Bertrand2009}.

In Section~\ref{seceval} we evaluate the MFT, discuss the significance
level in finite data sets and compare it to bootstrap methods. Most
importantly, we show by exemplary simulations that the MFA can have an
increased detection probability over single window techniques even when
a best window size is known. Thus, by using multiple window sizes, one
can detect rate changes in fast and slow time scales simultaneously,
increase the detection probability and avoid the problem of choosing
one near-optimal bandwidth
[cf., e.g., \citet{Basseville1993,Jones1996,Csorgo1998,Nawrot99,Shimazaki2007}].

Finally, we apply the MFT to a sample data set of spike train
recordings obtained as spontaneous single-unit activity from identified
dopamine neurons in the substantia nigra of anesthetized mice
(Section~\ref{sectdataanalysis}). In the sample data set, the detected
change points agree closely with visual inspection. In over 70\% of all
spike trains, which are classified to have a nonstationary rate,
different change points are detected by different window sizes.

\section{The point process model} \label{sectmodelass}
In this section we extend the assumptions of classical renewal
processes by introducing a class of renewal processes with varying
variance (RPVV) (Section~\ref{sectRPVV}). These processes are assumed
rate stationary, but the variance of life times may show a certain
degree of variability. Examples of such processes are given in
Section~\ref{examplegenren}. For the alternative hypothesis
(Section~\ref{sectfullmodel}) we combine several null elements,
resulting in processes with a piecewise stationary rate. In this model
we aim to detect rate changes irrespective of other point process
properties, such as the variability of the life times or even changes
in the variability of life times.

\subsection{Renewal processes with varying variance (RPVV)}\label{sectRPVV}
We write a point process $\Phi$ as an increasing sequence of events
\[
0 < S_1 < S_2 < S_3<\cdots,
\]
where $S_i$ denotes the occurrence time of the $i$th event, for $i=
1,2,\ldots.$ Alternatively, $\Phi$ is determined by its life times $(\xi
_i)_{i\ge1}$, where
\[
\xi_1 = S_1 \quad\mbox{and}\quad\xi_i =
S_i - S_{i-1}\qquad\mbox{for }i=2,3,\ldots,
\]
or by the counting process $(N_t)_{t\ge0}$, where
%
\begin{equation}\label{defnt} N_t=\max\{i\ge1 \mid S_i\le t\},\qquad t
\ge0,
\end{equation}
with the convention $\max\varnothing:= 0$.

Under the null hypothesis, we assume that a spike train can be
described as an element $\Phi$ of the following family of rate
stationary processes, which we term renewal processes with varying
variance (RPVV).

\begin{defi}[{[Renewal process with varying variance (RPVV)]}]\label{defgrp}
Let \mbox{$T>0$}, and let $\Phi$ be a renewal process restricted on $(0,T]$
whose life times, $\xi_1,\xi_2,\ldots,$ are assumed to be independent,
positive and square-integrable random variables with positive
variances, such that
for some $\mu,\sigma,c>0$ and all $\varepsilon>0$, with asymptotics as
$n\to\infty$, we have
%
\begin{eqnarray}
&& \mbox{rate stationarity:}\qquad\hspace*{57pt} \mathbb{E}[\xi_i] =\mu\qquad\mbox{for all } i\in\mathbb{N},\label{nullconda}
\\
&&\mbox{variance regularity:}\qquad\hspace*{42pt} \frac{1}{n}\sum_{i=1}^{n}\Var {(\xi_i)} \to\sigma^2,\label{nullcondd}
\\
&& \mbox{Lindeberg condition:}
\nonumber\\[-8pt]\label{nullcondb} \\[-8pt]\nonumber
&&\qquad \frac{ \sum_{i=1}^{n}\mathbb{E}[(\xi_i-\mu)^2 \mathbbm{1}_{\{(\xi_i-\mu)^2>\varepsilon^2 \sum_{i=1}^{n}\Var{(\xi_i)}\}}]} {\sum_{i=1}^{n}\Var{(\xi_i)}} \to 0,
\\
&& \mbox{uniform variance bound:}\qquad\hspace*{18pt} \sup_{i \in\mathbb{N}} \Var {(\xi_i)}< c, \label{nullcondc}
\\
&& \mbox{SLLN for squared life times:} \qquad \frac{1}{n}\sum_{i=1}^n \bigl(\xi _i^2-\mathbb{E}\bigl[\xi_i^2\bigr]\bigr)\to0\qquad \mbox{a.s.}\label{nullconde}
\end{eqnarray}
\end{defi}

Thus, an RPVV can be a renewal process with i.i.d.~life times, and thus
constant variance of life times. This applies, for example, to Poisson
processes or to processes with independent and $\Gamma(p,\lambda
)$-distributed life times, called here Gamma-processes. In addition,
the variance of life times can also show a certain variability as
specified in (\ref{nullcondd}) and (\ref{nullcondc}). Assumptions
(\ref{nullconda})--(\ref{nullconde}) are technically sufficient
for the asymptotic results that support our methods: condition (\ref{nullcondd}) imposes a regularity of the life times' variances over
time. The Lindeberg condition (\ref{nullcondb}) is later used for
process convergence to Brownian motion that allows to deduce
asymptotics for the related counting process. There, condition (\ref
{nullcondc}) will be used additionally. Assumption (\ref{nullconde}) is the strong law of large numbers (SLLN) for the
squares of the life times, which will be needed for strong consistency
of an estimation of $\sigma^2$ below. Note that by Kolmogorov's
conditions [\citet{Petrov1995}, Theorem~6.8] $(\xi_i^2)_{i\ge1}$
satisfying the SLLN is equivalent to
%
\begin{eqnarray}
\sum_{i=1}^{\infty}P\bigl(\bigl\llvert
\xi_i^2 - \mathbb{E}\bigl[\xi_i^2
\bigr]\bigr\rrvert \ge i\bigr) &<& \infty, \label{kol1}
\\
\sum_{i=1}^\infty\frac{1}{i^2} \mathbb{E}
\bigl[\bigl(\xi_i^{2}-\mathbb{E}\bigl[\xi
_i^{2}\bigr]\bigr)^2 \mathbbm{1}_{\{\llvert \xi_i^2-\mathbb{E}[\xi_i^{2}]\rrvert  <i\}}
\bigr] &<&\infty, \label{kol2}
\\
\frac{1}{n}\sum_{i=1}^n \mathbb{E}
\bigl[\bigl(\xi_i^2-\mathbb{E}\bigl[\xi_i^2
\bigr]\bigr)\mathbbm {1}_{\{\llvert \xi_i^2-\mathbb{E}[\xi_i^{2}]\rrvert  <n\}}\bigr] &\to& 0\qquad\mbox{as } n\to
\infty.\label{kol3}
\end{eqnarray}

The most important assumption (\ref{nullconda}) states that in an
RPVV, the mean rate $1/\mu$ is constant across time. We therefore also
use the short notation $\Phi(\mu)$.

%
\begin{figure}

\includegraphics{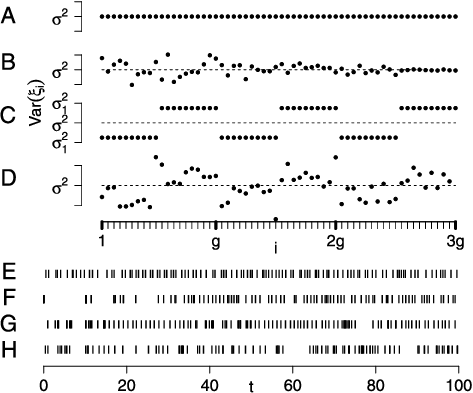}

\caption{Examples of RPVVs according to Definition \protect\ref{defgrp}. \textup{(A--D)} The variances $\Var(\xi_i)$ of life times
$\xi_i$ is indicated by points. $\Var(\xi_i)$ can be constant
$(\sigma^2)$ \textup{(A)}, can converge to a constant $\sigma^2$ \textup{(B)}, or can be
a step function alternating between different, fixed values in a
regular manner \textup{(C)}. In \textup{(D)}, the mean variance of the $g$ life times
$(\xi_1,\ldots,\xi_g)$, $(\xi_{g+1},\ldots,\xi_{2g})$, etc., is a
constant $\sigma^2$. \textup{(E--H)}~Realizations $(T=100)$ of point processes
with $\Gamma(p_i,\lambda_i)$-distributed life times $\xi_i$ with
constant expectation $\mathbb{E}[\xi_i] = p_i/\lambda_i=1$, that is,
$p_i = \lambda_i$. The variances $\Var(\xi_i)=p_i/\lambda_i^2
= 1/\lambda_i$ are given in \textup{(A--D)}, respectively. \textup{(E)} Independent and
$\Gamma(5,5)$-distributed life times with constant variance $\Var(\xi_i)= \sigma^2 = 1/5$.
\textup{(F)} $\Var(\xi_i)=1/\lambda_i\to
\sigma^2= 0.1$. \textup{(G)} The variance alternates in a regular manner,
changing after $g/2=20$ life times between $p_i=\lambda_i=1$ (Poisson
process) and $p_i=\lambda_i=20$ (a regular Gamma process). \textup{(H)} For
$g=40$, the mean variance of the $g$ life times $(\xi_1,\ldots,\xi_g)$,
$(\xi_{g+1},\ldots,\xi_{2g})$, etc., equals unity.} 
\label{bspgeneralizedrenewal}
\end{figure}

\subsection{Examples of RPVVs}\label{examplegenren}
Here, we give examples of point processes that satisfy the assumptions
of an RPVV from Definition~\ref{defgrp}. We assume rate stationarity
[condition (\ref{nullconda})]. Figure~\ref{bspgeneralizedrenewal}
shows examples of such processes. Panels A--D indicate the evolution of
variances of life times, and panels E--F illustrate point processes with
the corresponding variances and Gamma-distributed life times. Because
Gamma-processes have been used frequently in order to describe neuronal
spiking activity
[cf., and the references therein, \citet{Nawrot2008}], we also
use Gamma processes for all simulations in the present article,
choosing suitable combinations of rate and regularity parameters for
each simulation.

The most simple example of an RPVV is a process with i.i.d.~life times
(Figure~\ref{bspgeneralizedrenewal}A~and~E). As a second example
(Figure~\ref{bspgeneralizedrenewal}B~and~F), an RPVV can be a
process in which the variances of life times converge to a constant.
Third, the variance of life times can alter regularly between two
different values (Figure~\ref{bspgeneralizedrenewal}C). The
corresponding point process (panel~G) shows regular and irregular
sections. This example can be extended such that the mean variance of
life times is constant at equidistant grid points $g, 2g,\ldots$
(Figure~\ref{bspgeneralizedrenewal}D~and~H).

\subsection{The full model}\label{sectfullmodel}
In contrast to the null assumption, the alternative hypothesis assumes
that $\Phi$ is piecewise an RPVV, where the mean rate can change
between the different sections. Formally, we assume that under the
alternative hypothesis, a spike train is an element of the class
constructed in~Construction~\ref{constructionrenewalmodel}.

%
\begin{cons}\label{constructionrenewalmodel}
Let $T>0$, and let $\mathbf{C}$ denote the set of all finite subsets of
$(0,T]$. Assume $C:=\{c_1,\ldots,c_k\}\in\mathbf{C}$, with $c_1<\cdots<c_k$.

At time $0$ start $k+1$ independent RPVVs $\Phi_1(\mu
_1),\ldots,\Phi_{k+1}(\mu_{k+1})$ with
\[
\mu_i \neq\mu_{i+1}\qquad\mbox{for } i=1,\ldots,k.
\]
Let $c_0:=0$, $c_{k+1}:=T$ and define
%
\begin{equation}
\label{gammamodel} \Phi:= \bigcup_{i=1}^{k+1}
\Phi_i |_{(c_{i-1},c_i]},
\end{equation}
where $\Phi_i \mid_{(c_{i-1},c_i]}$ denotes the restriction of $\Phi_i$ to
the interval $(c_{i-1},c_i]$.
\end{cons}

\begin{figure}

\includegraphics{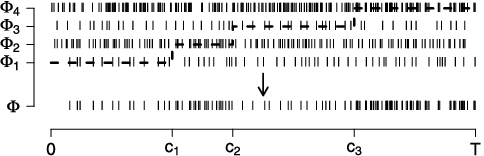}

\caption{The change point model combines a set of RPVVs. A realization
of a process $\Phi$ on $(0,T]$ that results from Construction~\protect\ref{constructionrenewalmodel}. $\Phi$ has three change points
$c_1,c_2,c_3$ and originates from the four RPVVs $\Phi_1,\ldots,\Phi
_4$, jumping from process $\Phi_i$ to $\Phi_{i+1}$ at change point $c_i$.}
\label{renewalmodel}
\end{figure}

The times $c_1,\ldots,c_k$ are called change points. An example of a
point process generated according to this construction is shown in
Figure~\ref{renewalmodel}. The resulting rate of $\Phi$ is a step
function with change points $c_1,\ldots,c_k$.

%

We now define a model set
$\mathscr{M}:=\mathscr{M}(T)$ to be the family of processes that derive
from Construction~\ref{constructionrenewalmodel} and test the null hypothesis:
\begin{longlist}[$H_A$:]
\item[$H_0$:]
$\Phi\in\mathscr{M}$ with $C=\varnothing$, that is, $\Phi$ is an RPVV,
in particular rate stationary,
against the alternative.
\item[$H_A$:]
$\Phi\in\mathscr{M}$ and $C\neq\varnothing$, that is, there is at least
one change point.
\end{longlist}

\section{The multiple filter test (MFT)}\label{secmft}
In order to test the above null hypothesis of rate stationarity in the
model set $\mathscr{M}$, we derive here a multiple filter test (MFT).
Section~\ref{sectmft} summarizes the construction of the test. Details
on parameter estimation and limit results are given in Sections~\ref{varianceestimation} and~\ref{subseclimit}.

\subsection{Derivation of the MFT}\label{sectmft}
\begin{figure}

\includegraphics{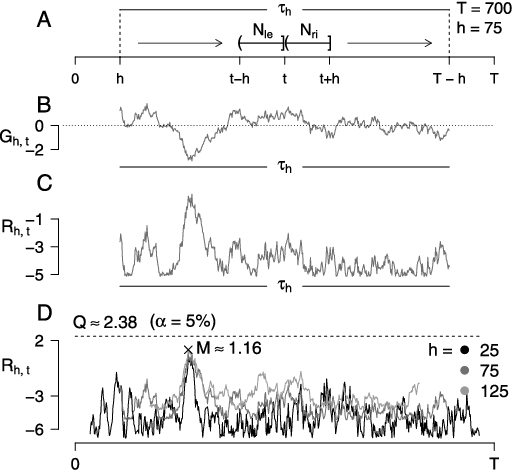}

\caption{Illustration of the computational steps and processes
involved in the MFT. The MFT is applied here to a stationary process on
$(0,700]$ with independent and $\Gamma(0.25,5)$-distributed life times.
\textup{(A)} For one window size $h=75$, the number of events in the left and
right window, $N_{\ler},N_{\ri}$, are derived for every $t\in\tau_h$. \textup{(B)}
The process $(G_{h,t})_{t\in\tau_h}$ for one window $h=75$. \textup{(C)} The
scaled process $(R_{h,t})_{t\in\tau_h}$ for one window $h=75$. \textup{(D)} All
scaled processes $(R_{h,t})_{t\in\tau_h}$ for $h\in H=\{25,75,125\}$.
Different gray shades indicate different window sizes, the asymptotic
threshold $Q$ is represented by a dashed line. Here, the test statistic
$M=\max_{h,t}R_{h,t}<Q$ and, thus, the null hypothesis of rate
stationarity is not rejected.}
\label{ghtschematisch}
\end{figure}

The main idea of the MFT is to extend a filtered derivative technique
[see the contributions \citet
{Basseville1993,Brodsky1993,Csorgo1998}], which slides two adjacent
windows of size $h$ and compares the number of events in the left and
right window. Formally, let $T>0$ and $\Phi$ be an element of the model
set $\mathscr M$. For $h\in(0,T/2]$ we define an analysis region $\tau
_h:= (h,T-h]$. Let $N_{(a,b]}(\Phi)$ denote the number of elements of
$\Phi$ in the interval $(a,b]\subset(0,T]$.
For each point $t\in\tau_h$ we compare the number of events
\[
N_{\ler}:= N_{(t-h,t]}(\Phi)\quad\mbox{and}\quad N_{\ri}:= N_{(t,t+h]}(\Phi)
\]
in the left and right window (Figure~\ref{ghtschematisch}A).

%

A large difference $N_{\ri}-N_{\ler}$ can indicate deviations from the
null hypothesis of rate stationarity. But because the variance of the
difference depends on process parameters, the difference
$N_{\ri}-N_{\ler}$ will be normed as follows:
%
\begin{equation}
\label{auxiliaryvariable} G_{h,t}:= G_{h,t}(\Phi):= \frac{N_{\ri}-N_{\ler}}{\hat s}
\qquad\mbox{if }\hat s > 0,
\end{equation}
and $G_{h,t}:=0$ if $\hat s = 0$
for all $t\in\tau_h$ (Figure~\ref{ghtschematisch}B). The term $\hat s$ denotes an estimator of $\sqrt{\Var[N_{\ri}-N_{\ler}]}$,
which is defined in (\ref{schaetzers}). We will show in Section~\ref
{subseclimit} that the process $(G_{h,t})_{t\in\tau_h}$ converges to a
$2h$-dependent Gaussian process $(L_{h,t})_{t\in\tau_h}$. The limit
process $(L_{h,t})_{t\in\tau_h}$ is a continuous functional of a
standard Brownian motion and depends only on $T$ and $h$. In
particular, it is independent from the parameters of $\Phi$ such as,
for example, the rate or regularity.


Large absolute values of $G_{h,t}$ indicate potential deviations from
rate stationarity. Therefore, the maximum
\[
M_h:=\max_{t\in\tau_h} \llvert G_{h,t}\rrvert
\]
can serve as a test statistic for a single window.

In order to combine multiple window sizes of a finite set $H\subset
(0,T/2]$, we consider a set of stochastic processes $\{(G_{h,t})_{t\in
\tau_h}\mid h\in H\}$, which are all derived from the same underlying point
process $\Phi$. Each process $(G_{h,t})_{t\in\tau_h}$ results in one
maximum $M_h$. Instead of using the raw maxima $M_h$, we suggest to
standardize $M_h$ because the distribution of $M_h$ depends on $h$. As
mentioned above, the process $(G_{h,t})_{t\in\tau_h}$ is
$2h$-dependent, and a smaller $h$ results in weaker temporal
dependencies of the process. This leads to higher chance fluctuations
in $(G_{h,t})_t$ for smaller~$h$, and thus a higher rejection threshold.

If the expectation and variance of $M_h$ were known, we could use the term
%
\begin{equation}
\label{supweighted} \frac{M_h-\mathbb{E}[M_h]}{\sqrt{\Var(M_h)}}
\end{equation}
in order to give every window a similar impact on the global maximum of
all processes. Here, we approximate the expectation and variance using
simulations of the set of limit processes $\{(L_{h,t})_{t\in\tau
_h}\mid h\in H\}$. Defining $M_h^*:=\sup_{t\in\tau_h}\llvert  L_{h,t}\rrvert  $, we
approximate the expectation $\mathbb{E}[M_h]$ by the empirical mean
$\overline M_h^*$ and the variance $\Var(M_h)$ by the
empirical variance $v(M_h^*)$. The resulting test statistic $M$ across
all windows is defined as the global maximum
%
\begin{equation}
\label{teststatistic} M:= \max_{h\in H} \biggl(\frac{ M_h - \overline M_h^*}{\sqrt
{v(M_h^*)}}
\biggr).
\end{equation}

Finally, we reject the null hypothesis at level $\alpha$ if
$M>Q:=Q(\alpha,T,H)$. The threshold $Q$ is defined such that under the
null hypothesis, $M>Q$ with probability~$\alpha$. In order to derive
$Q$, one can again use the limit processes $\{(L_{h,t})_{t\in\tau
_h}\mid  h\in H\}$ and approximate $Q$ by the empirical quantile of
%
\begin{equation}
\label{supremelimit} M^*:=\sup_{h \in H} \biggl(\frac{ M_h^* - \overline M_h^*}{\sqrt
{v(M_h^*)}}
\biggr).
\end{equation}
Note that all limit processes $(L_{h,t})_t$ are derived from the same
Brownian motion in order to ensure comparability with the processes
$(G_{h,t})_t$, which result from the same point process $\Phi$.

For change point detection explained later in Section~\ref{seccpd} and
for graphical illustration, we use the scaled process
%
\begin{equation}
\label{Rht} R_{h,t}:= \biggl(\frac{ \llvert  G_{h,t}\rrvert   - \overline M_h^*}{\sqrt
{v(M_h^*)}} \biggr)\qquad
\mbox{(Figure~\ref{ghtschematisch}C),}
\end{equation}
which scales $(G_{h,t})_{t\in\tau_h}$ and accounts for the scaling of
the maxima. Because the maximum of all processes $(R_{h,t})$,
%
\begin{equation}
M= \max_{h\in H}\max_{t\in\tau_h} R_{h,t}
\end{equation}
is identical to the above global test statistic, it can be read
directly from the graph. The processes $(R_{h,t})_{t\in\tau_h}$ and
their comparison with the threshold $Q$ are illustrated in Figure~\ref{ghtschematisch}D.

\subsection{Variance estimation}\label{varianceestimation}

By definition of our auxiliary variables $G_{h,t}$ [see~(\ref
{auxiliaryvariable})], we need to specify an estimator $\hat s^2$ for
the variance of $N_{\ri} - N_{\ler}$. The idea is to estimate the variance
from the life times of the elements in the left and right windows of $G_{h,t}$.

Let $\xi_1,\xi_2,\ldots$ be the life times of an RPVV with constant $\mu
$ and $\sigma^2$ as in (\ref{nullconda})~and~(\ref{nullcondd}).
Given $T$ and $h$, for every $t\in\tau_h$ we define
%
\begin{equation}
\gamma_{\ler}(t,h):= \bigl\{\xi_i\dvtx  S_i,S_{i-1}
\in(t-h,t], i=1,2,\ldots\bigr\},
\end{equation}
the set of all life times that correspond to events in the left window.
We relabel this set of life times $\xi_1^{\ler},\xi_2^{\ler},\ldots.$
Analogously for the right window, we obtain $\gamma_{\ri}(t,h) = \{\xi
_1^{\ri},\xi_2^{\ri},\ldots\}$.

The empirical mean of the life times in the left window is denoted by
%
\begin{equation}
\label{muschaetzer1} \hat\mu_{\ler}:= \hat\mu_{\ler}(t,h):=\overline{\gamma_{\ler}(t,h)}\qquad \mbox{if } \llvert \gamma_{\ler}\rrvert
> 0,
\end{equation}
and $\hat\mu_{\ler}:= 0$ if $\llvert  \gamma_{\ler}\rrvert   = 0$. The empirical variance
of the life times is
%
\begin{equation}
\label{sigmaschaetzer1} \hat\sigma_{\ler}^2:= \hat\sigma_{\ler}^2(t,h):=v\bigl(\gamma_{\ler}(t,h)\bigr)
\qquad \mbox{if } \llvert \gamma_{\ler}\rrvert > 1,
\end{equation}
and $\hat\sigma_{\ler}^2:=0$ if $\llvert  \gamma_{\ler}\rrvert   \le1$. The bar denotes
the empirical mean, $v(\cdot)$ denotes the corrected sample variance of
$\gamma_{\ler}(t,h)$, and $\llvert  \cdot\rrvert  $ denotes the number of elements.
Analogously, we define $\hat\mu_{\ri}$ and $\hat\sigma_{\ri}^2$ for the
right window.

As an estimator for the variance of $N_{\ri}-N_{\ler}$ we propose
%
\begin{equation}
\label{schaetzers} \hat s^2:= \hat s^2(t,h):= \biggl(
\frac{\hat\sigma_{\ri}^2}{\hat\mu
_{\ri}^3} + \frac{\hat\sigma_{\ler}^2}{\hat\mu_{\ler}^3} \biggr) h\qquad\mbox {if } \hat\mu_{\ler} \wedge\hat\mu_{\ri} > 0,
\end{equation}
and $\hat s^2:=0$ otherwise,
where $\wedge$ denotes the minimum. Note that $\hat s^2$ is zero by
definition if the number of events is less than two in any window. We
prove strong consistency of these estimators in an appropriate
asymptotic setting in Appendix~\ref{convesti}. Heuristically, this
estimator is suggested by the fact that under our conditions on the
life times of the RPVV we obtain for the number
$N_t$ of events up to time $t$ that, as $t \to\infty$, we have
%
\begin{equation}
\label{ntgrenzwertallgemein} \frac{N_t- \nicefrac{t}{\mu}}{\sqrt{ \nicefrac{t \sigma^2}{\mu^3}}} \stackrel{d} {\longrightarrow} N(0,1)\quad
\mbox{and}\quad\Var [N_t] \sim\sigma^2t/
\mu^3,
\end{equation}
where $\stackrel{d}{\longrightarrow}$ denotes convergence in
distribution. Hence, we obtain
\[
\Var[N_{\ri}-N_{\ler}]\approx \biggl(
\frac{\sigma^2}{\mu^3} + \frac{\sigma^2}{\mu^3} \biggr) h.
\]

\subsection{Limit distribution of $(G_{h,t})$ under $H_0$}\label{subseclimit}
In order to compute the test statistic $M$ and choose the rejection
threshold $Q$, we derive a limit of the process $(G_{h,t})_{t\in\tau
_h}$, choosing an asymptotic setting in which time $T$ and window size
$h$ grow proportionally.
As the limit we identify a $2h$-dependent Gaussian process
$(L_{h,t})_{t\in\tau_h} $ on $\tau_h$ that does not depend on the
parameters of the process $\Phi$.

To make this asymptotic statement precise, let $\Phi$ be an element of
$H_0$ with life times $\xi_1,\xi_2,\ldots.$ We consider an extended
version $(G_{h,t}^{(n)})_{t\in\tau_h}$ of $(G_{h,t})_{t\in\tau_h}$,
%
\begin{equation}
\label{ghtextension} G_{h,t}^{(n)}:=\frac{(N_{n(t+h)}-N_{nt})-(N_{nt}-N_{n(t-h)})}{\hat s(nt,nh)}\qquad\mbox{if
}\hat s(nt,nh) >0,
\end{equation}
and $G_{h,t}^{(n)}:= 0$ otherwise, for all $t\ge h$ and $n=1,2,\ldots.$
Recall that $N_t$ denotes the number of life times up to time $t$ and
the estimator $\hat s$ is defined in (\ref{schaetzers}). We consider
the processes $(G_{h,t})_{t\in\tau_h}$ and $(G_{h,t}^{(n)})_{t\in\tau
_h}$ as c\`adl\`ag processes in the Skorokhod topology.

\begin{figure}

\includegraphics{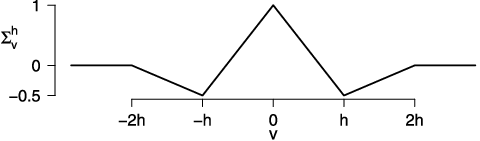}

\caption{The autocovariance structure $\Sigma_{v}^h$ of
$(L_{h,t})_{t\ge h}$ as a function of the time lag $v$ for a fixed
window size $h$.}
\label{kovarianzplot}
\end{figure}

The asymptotic analysis is given by letting $n\to\infty$.
To define the limit process, let $W = (W_t)_{t\ge0}$ denote a standard
Brownian motion on $[0,\infty)$. For $h>0$ we define for all $t\ge h$
%
\begin{equation}
\label{limitprocess} L_{h,t}:= \frac{(W_{t+h} - W_t) - (W_t - W_{t-h})}{\sqrt{2h}}.
\end{equation}
The process $(L_{h,t})_{t\ge h}$ is a $2h$-dependent Gaussian process,
with zero mean and autocovariance given as
%
\begin{eqnarray}
\label{sigmahomogen}
\Sigma_{v}^h:= \Sigma_{u,u+v}^h:= \cases{ \displaystyle1-\frac{3}{2h}\llvert v\rrvert, &\quad if $\llvert
v\rrvert \in[0,h]$,
\cr
\displaystyle-1 +\frac{1}{2h}\llvert v\rrvert, &
\quad if $\llvert v\rrvert \in(h,2h]$,
\cr
0, &\quad if $\llvert v\rrvert \ge2h$,}
\end{eqnarray}
for all suitable $u,v$ (Figure~\ref{kovarianzplot}). Note that the
autocovariance only depends on the window size $h$ and the time lag $v$
of two elements $L_{h,t}$ and $L_{h,t+v}$.

%
In the \hyperref[app]{Appendix} we show the following process convergence, extending
results obtained by \citet{Steinebach1995}.

%
\begin{theo}\label{hauptaussagegeneralisiert}
Let $T>0$ and $h\in(0,T/2]$ be a window size. Let $\Phi$ be an element
of the null hypothesis.
Then for the processes $(G_{h,t}^{(n)})$ and $(L_{h,t})$ defined in~(\ref{ghtextension}), and (\ref{limitprocess}), as $n\to\infty$, we have
%
\begin{equation}
\label{centconv} \bigl(G_{h,t}^{(n)}\bigr)_{t\in\tau_h}
\stackrel{d} {\longrightarrow} (L_{h,t})_{t\in\tau_h},
\end{equation}
where $\stackrel{d}{\longrightarrow}$ denotes weak convergence in the
Skorokhod topology.
\end{theo}

\section{Multiple filter algorithm (MFA) for change point detection}
\label{seccpd}
In Section~\ref{secmft} the first part of the MFT was presented as a
test for the null hypothesis of rate stationarity versus the
alternative of at least one rate change. After rejection of the null
hypothesis, we intend to identify the number and location of change
points. To this end, we propose an algorithm that combines the results
of multiple window sizes. It consists of a procedure for change point
detection on the basis of individual windows (single filter algorithm---SFA, Section~\ref{sectSWD}) and a multiple filter algorithm (MFA) for
the combination of individual windows (Section~\ref{sectMFA}).

\subsection{Single filter algorithm (SFA)}\label{sectSWD}
For the detection of change points with a single window of size $h$, we
apply a common method to the scaled filtered derivative process
$(R_{h,t})_{t\in\tau_h}$, which successively estimates change points
from the maxima of the process [see the contributions \citet
{Basseville1993,Bertrand2000,Bertrand2009,Antoch1995}]. Similar
procedures have been shown to give consistent estimates of the number
and location of the change points under mild conditions in Gaussian
sequence change point models [\citet{Huskova2001,Muhsal2013}].

%
\begin{figure}

\includegraphics{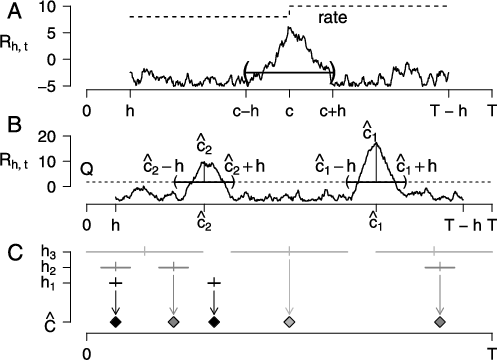}

\caption{The SFA and MFA. \textup{(A)}~A change point at time $c$ affects the
process $(R_{h,t})_t$ within the $h$-neighborhood of $c$, and the
maximum of $(R_{h,t})_t$ is expected at $c$. \textup{(B)} The SFA successively
searches for maxima of $(R_{h,t})_t$. When the maximum is larger than
$Q$, the maximizer $\hat c_1$ is the first change point estimator. Then
the $h$-neighborhood of $\hat c_1$ is omitted, and the procedure is
iterated on the remaining process until the maximum remains smaller
than $Q$ (underlying process different from~\textup{A}). \textup{(C)}~Schematic
representation of the MFA for a window set $H=\{h_1,h_2, h_3\}$
(underlying process different from \textup{A}~and~\textup{B}). The change points detected
by SFA are marked as vertical bars, and their $h_i$-neighborhoods are
indicated by horizontal lines ($h_1$, black, $h_2$, gray, $h_3$, light
gray). The MFA first accepts all change points detected with the
smallest window $h_1$ (black diamonds). Among the change points
detected by $h_2$ (gray), only the first one is rejected because its
$h_2$-neighborhood contains an accepted change point. Among the change
points detected by the largest window (light gray), only the second one
is added to the list of accepted change points because its
$h_3$-neighborhood does not contain formerly accepted change points.
Diamonds indicate finally accepted change points.}\label{owdschematisch}
\end{figure}

The SFA for one $h\in H$ works as follows. First, observe the maximum
of the process $(R_{h,t})_{t\in\tau_h}$. If $\max_t R_{h,t}>Q$, this
indicates deviations from rate stationarity. The time $\hat c_1$ at
which this maximum is taken is an estimate of a change point because
the maxima are expected at the change points if the difference between
change points is sufficiently large (Figure~\ref{owdschematisch}A).
More precisely, one should note that the sample path of $(R_{h,t})_{t\in
\tau_h}$ is a step function, so that the set of maximizers is an
interval. We define $\hat c_1$ as the infimum of this interval
\[
\hat c_1:= \inf\Bigl\{\arg\max_{t \in\tau_h}
R_{h,t}\Bigr\}.
\]
Second, we observe that a change point which occurs at time $c$ affects
the behavior of the process $(R_{h,t})_{t\in\tau_h}$ within the
$h$-neighborhood of $c$,
%
\begin{equation}
B_{h}(c):=(c-h, c+h) \cap\tau_h\qquad
\mbox{(Figure~\ref{owdschematisch}A),}
\end{equation}
while leaving all points outside of $B_{h}(c)$ unaffected. Therefore,
the $h$-neighbor\-hood of $\hat c_1$ is omitted in the subsequent
analysis. If the remaining process $(R_{h,t})_{t\in\tau_h\setminus
B_{h}(c)}$ outside of $B_{h}(c)$ exceeds $Q$, this indicates another
deviation from rate stationarity because a change point at $c$ cannot
cause this deviation. Therefore, we successively identify change points
as the maxima of $(R_{h,t})_t$ outside the union of all $B_{h}(\hat c_i)$ of detected change points, until the process $(R_{h,t})_t$ is
smaller than $Q$ in all remaining intervals (Figure~\ref{owdschematisch}B).

\subsection{Multiple filter algorithm (MFA)}\label{sectMFA}
We now propose a multiple filter algorithm with which the results of
the SFA of different windows can be combined. This integrates the
advantages of multiple time scales because large windows are more
likely to detect small rate changes and small windows can be more
sensitive to fast changes. In particular, using only a large window of
size $h$, the SFA can fail or mislocate change points $c_1,c_2$ with
distance smaller than $h$. This suggests to prefer change point
estimates of smaller windows.

The MFA can be summarized as follows (Figure~\ref{owdschematisch}C).
Let $H=\{h_1,h_2,\ldots,\break h_n\}$ be the set of involved windows, with
$h_1<\cdots<h_n$. Derive the threshold $Q$ for this set $H$ as
described in Section~\ref{secmft}. For all $h_i$, detect change points
via SFA. Let $\widehat C_i:=\{\hat c_1^i,\ldots, \hat c_{k_i}^i \}$
denote the set of change points estimated with window $h_i$. Then,
define a set of accepted change points $\widehat C$, which is first set
to $\widehat C:=\widehat C_1$, that is, all change points estimated by
the smallest window. Among the change points $\widehat C_2$ associated
with $h_2$, only those are added to $\widehat C$ whose
$h_2$-neighborhood does not include a formerly accepted change point
$c_j^1\in\widehat C$. The remaining estimates $c_j^2\in\widehat C_2$
are assumed to be affected by change points that have already been
estimated and therefore omitted in the further analysis. This procedure
is iterated by successively increasing the window sizes up to $h_n$.

\subsection{Application to a simulated point process}\label{subsecapplicatonofcpd}

\begin{figure}

\includegraphics{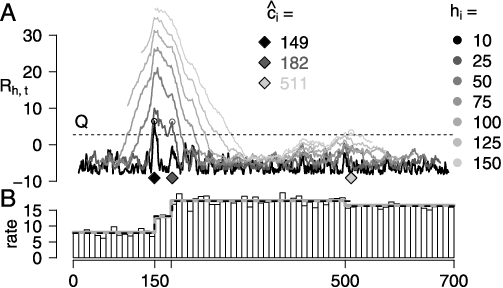}

\caption{Application of the MFT to a simulated point process on
$(0,700]$ with three change points at $c_1 = 150$, $c_2=180$ and
$c_3=500$. The life time distributions were $\exp(8)$ in $[0,c_1]$,
$\Gamma(2,26)$ in $[c_1,c_2]$, $\exp(18)$ in $[c_2,c_3]$ and $\Gamma
(2,33)$ in $[c_3,T]$, corresponding to rates of $8,13,18$ and $16.5$ in
the respective intervals. \textup{(A)} Gray curves indicate the processes
$(R_{h,t})$ for window sizes $h\in H=\{10,25,50,75,100,125,150\}$. The
simulated threshold $Q=2.75$ is indicated by the dashed line. The
estimated change points $\hat c_1 = 149, \hat c_2=182, \hat c_3 = 511$
are marked by diamonds. As indicated by the grayscale of the diamonds,
each change point was detected by a different window size. \textup{(B)} Rate
histogram of the underlying point process with real (gray) and
estimated (dashed) rate profiles.}
\label{cpdbsp}
\end{figure}

Figure~\ref{cpdbsp} illustrates the application of the MFA to a
simulated point process with three change points. All change points are
detected by the MFA, and the estimated change points correspond closely
to the true change points. Consequently, the rate estimates agree
closely with the true rates.

Figure~\ref{cpdbsp} also shows that different window sizes were used
for the detection of different change points: while the first change
point was detected by the smallest window $h_1=10$, the second was
detected by $h_2=25$ and the third by $h_3=150$. This supports the idea
of combining several windows: if change points are close together
(e.g., $c_1$ and $c_2$ in Figure~\ref{cpdbsp}), small windows are
preferable because large windows tend to be affected by both change
points, and thus lead to imprecise estimates. On the other hand, small
rate changes require large windows, which have a higher test power.
Indeed, none of the individual windows could detect all change points
(data not shown).

%

\subsection{Choosing the window set $H$}\label{sectwindowchoice}
The previous example and the simulations that follow in Section~\ref
{mfitp} show that multiple filters can increase the probability to
detect change points. This is because rate changes in fast and slow
time scales can be detected simultaneously using multiple windows.
However, using too many windows increases the threshold $Q$ applied for
change point detection, which can also decrease the test power in
certain settings. Therefore, we discuss here in which way $Q$ depends
on the window set $H$ and give recommendations for the choice of~$H$.

%
\begin{figure}

\includegraphics{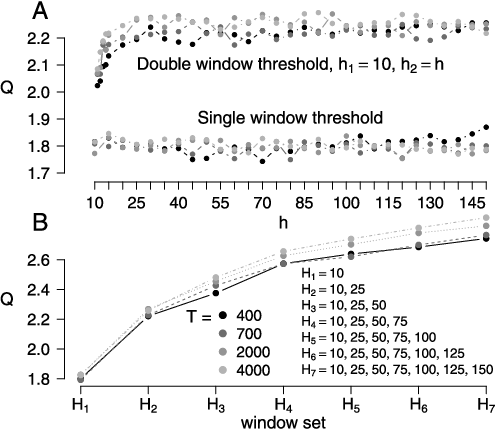}

\caption{Dependence of $Q$ on $H$ and $T$ for $\alpha=5\%$. \textup{(A)} For
SFA, the threshold $Q$ basically does not depend on $T$ or $h$
(connected points at about $Q\approx1.8$, color codes for $T$ as in~\textup{B}). Choosing $h_1=10$, the second window $h_2$ increases $Q$ to about
$2.23$, with a stronger increase for larger $h_2$ (points at about
$Q=2.2$, color codes for $T$ as in \textup{B}). \textup{(B)} Adding more windows from the
set $H=\{10,25,50,75,100,125,150\}$ leads to only slight increases in
$Q$. Dependence on the recording time $T$ is weak. 10,000 simulations
were used to calculate the empirical mean and standard deviation for
the standardization of the limit process $(L_{h,t})_t$ [equation
(\protect\ref{supremelimit})].}
\label{qdependence}
\end{figure}

Because $Q$ depends only on $T$, $H$ and $\alpha$, we investigate its
dependency on $T$ and $H$ for $\alpha=5\%$. Figure~\ref{qdependence}A
shows that if only one window is used, the single window threshold $Q$
does essentially not depend on $h$ or $T$. Because the test statistic
is normed for every $h$ [equation (\ref{supweighted})], any window
size $h$ and any simulated time $T$ results in a threshold of about
$Q\approx1.8$. In order to study the influence of one additional window
on $Q$, we fix $h_1=10$ and illustrate the double window threshold $Q$
for $h_2\in\{11,12,\ldots,15,20,\ldots,150\}$ in Figure~\ref{qdependence}A. The threshold $Q$ increases to about $2.23$. Smaller
$h_2$ close to $h_1=10$ lead to smaller increases than larger windows
because the processes $(R_{h_1,t})_t$ and $(R_{h_2,t})_t$ show higher
correlation if $\llvert  h_1-h_2\rrvert  $ is small. In Figure~\ref{qdependence}B we
successively add windows of increasing size to the set $H_7=\{
10,25,50,75,100,125,150\}$. The increase in $Q$ from $H_1=\{h_1\}$ to
$H_2=\{h_1,h_2\}$ is about the same as from $H_2$ to $H_7$. Similarly,
adding more windows between $10$ and $150$ would only slightly increase
$Q$ (data not shown).

Because additional windows have minor impact on $Q$, we recommend the
following window choice: the smallest window $h_1$ should be restricted
such that the asymptotic significance level is approximately kept. To
this end, Section~\ref{sectsignlevel} investigates the empirical
significance level for stationary Gamma processes with different
regularity and rate parameters. The maximal window $h_{\max}$ is only
limited by $T/2$. The choice of the grid between $h_1$ and $h_{\max}$
can be guided by the following principles: choosing a narrow grid can
detect change points in a broad class of time scales. However, it will
also slightly increase the threshold $Q$, and thus reduce the
probability to detect change points at all. Additionally, it increases
the computational effort required for the performance of the test.
Here, we study the performance for the window set $H=\{
10,25,50,75,100,125,150\}$.

\section{Evaluation of the MFT}\label{seceval}

\subsection{Practical applicability of the MFT}
\subsubsection{Empirical significance level in simulations} \label{sectsignlevel}
As discussed in Section~\ref{subseclimit}, the proposed MFT is an
asymptotic procedure, providing asymptotic significance level $\alpha$.
Therefore, we use simulations in order to investigate under which
conditions the asymptotic significance level is kept for small rates in
the finite setting. We simulate rate stationary renewal processes with
Gamma-distributed life times in order to investigate the empirical
significance level of the asymptotic MFT. We focus on the parameters
$T=700$, $H=\{10,25,50,75,100,125,150\}$ and an asymptotic significance
level $\alpha=5\%$.

%
\begin{figure}[b]

\includegraphics{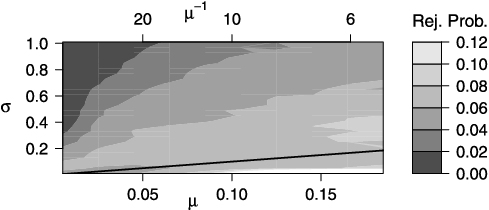}

\caption{Simulated rejection probability of the MFT for processes with
i.i.d.~Gamma-distributed life times ($T=700$, $H=\{
10,25,50,75,100,125,150\}$, 10,000 simulations). For high
irregularity, the test tends to be conservative. With increasing
regularity, the rate required to keep the asymptotic $5\%$ significance
level increases. ${\mu}$ and $\sigma$ denote the mean and standard
deviation of life times.}
\label{rejprobgamma}
\end{figure}

Figure~\ref{rejprobgamma} shows the empirical significance level
obtained in 10,000 simulations as a function of the mean $(\mu)$ and
standard deviation $(\sigma)$ of the independent and Gamma-distributed
life times. Under high irregularity, that is, if $\sigma$ is high, the
test remains conservative. With increasing regularity, the rate
required to obtain an empirical significance level of $5\%$ is
increasing. For low rates and high regularity, the percentage of false
positives of the MFT tends to be slightly larger than the asymptotic
significance level. In the very extreme case of almost perfect
regularity and low rates (white area in the bottom right corner), the
MFT should not be applied because the empirical significance level is
largely enhanced. In all but these extreme parameter combinations, the
detection of more than one change point was very unlikely under the
null hypothesis (detection of at least $2$ change points in $<$1\%, of
at least $3$ change points in $<$0.1\% of $1000$ simulations, data not
shown). Thus, the detection of more than one change point can almost
always be considered a strong indication of rate nonstationarity.

In summary, one needs to keep in mind for practical applications that
the error rate can be slightly enhanced for regular processes with low
rates. However, a false estimation of a nonexisting change point is not
problematic if one primarily intends to split up the time series into
rate stationary sections. If the significance level needs to be kept
strictly even for small rates, the window size needs to be increased.
This has the same effect as increasing the rate because the
approximation of $(G_{h,t})_t$ to the limit process $(L_{h,t})_t$
[equation (\ref{hauptaussagegeneralisiert})] mainly depends on the
mean number of events per window.

\subsubsection{Comparison of the MFT to a bootstrap test}\label{sectboot}
The preceding section shows that the MFT should be treated carefully in
situations with limited rates and high regularity because the
asymptotic significance level is not precisely kept. Therefore, one
might consider deriving $Q$ with a bootstrap procedure, as suggested,
for example, by \citet{Huskova2001}. The distribution of $M$ can
then be derived directly by permutation of the life times and
recalculation of $M$ in the permuted process. By construction, this
procedure yields an empirical significance level of $5\%$ if the
underlying process is a classical, stationary renewal process. However,
it has two shortcomings: first, it requires high computational effort
because the process $(R_{h,t})_t$ [equation (\ref{Rht})] needs to be
recalculated for every realization. Second, permutation can only be
applied if the life times are independent and identically distributed.

Therefore, we compare the MFT with a bootstrap test when the underlying
process does not comply with the assumption of independent and
identically distributed life times, that is, when the underlying
process is a rate stationary RPVV but not a classical renewal process.
To this end, we simulate rate stationary processes with
Gamma-distributed life times. The variance of life times changes every
$g/2$ life times, alternating between two values. As shown in
Section~\ref{examplegenren} (Figure~\ref{bspgeneralizedrenewal}C),
the resulting process is an RPVV. 

In order to reduce computational effort for the bootstrap test, we
replace $R_{h,t}$ by only computing
$\llvert  N_{\ri}(t,h)-N_{\ler}(t,h)\rrvert  $,
the absolute difference of the number of events in the left and right
windows, for every $h$ and $t$, and derive the maximum of these values
as a test statistic. The $95\%$-quantile of the distribution of this
test statistic is then estimated in permutations, and the null
hypothesis is rejected if the maximum is larger than its estimated quantile.

Table~\ref{mftvsboot} shows the resulting significance levels for the
MFT and the bootstrap procedure. The MFT roughly keeps the $5\%$
significance level in all simulated scenarios, whereas the bootstrap
test rejects the null hypothesis in about $3\%$, $7\%$ and $15\%$ of
the simulations. This indicates, as expected, that permutation tests
are not necessarily robust against changes in the variance of life
times and should therefore not be applied under such conditions.

%
\begin{table}
\tabcolsep=0pt
\tablewidth=250pt
\caption{Comparison of the significance level of the MFT and a
bootstrap test for simulated RPVVs. Here, the distribution of life
times changes every $g/2$ life times from $\Gamma(0.5,15)$ to $\Gamma
(5,150)$, leading to alternations between regular and irregular
patterns. The grid size is~\textup{(A)}~$g=5000$, (\textup{B}) $g={}$10,000, (\textup{C}) $g={}$20,000.
$1000$ simulations with $H=\{10,25,50,75,100,125,150\}$ and $T=700$ at
level $\alpha=5\%$ were performed in all cases, $1000$ permutations
were used for the construction of the bootstrap threshold}\label{mftvsboot}
\begin{tabular*}{\tablewidth}{@{\extracolsep{\fill}}@{}lc c @{}}
\hline
$\bolds{\Gamma(0.5,15) \leadsto\Gamma(5,150)}$&  \textbf{MFT} & \textbf{Bootstrap}\\
\hline
(A) $g=5000$ & $(5.9 \pm 0.7)\%$ & $(3.0 \pm 0.5)\%$\\
(B) $g={}$10,000 & $(4.7 \pm 0.7)\%$ & $(6.6 \pm 0.8)\%$ \\
(C) $g={}$20,000 & $(5.5 \pm 0.7)\%$ & $(15.1 \pm1.1)\%$ \\
\hline
\end{tabular*}
\end{table}

\subsubsection{True change points do not increase the frequency of
falsely detected change points}
The previous paragraphs show that the proposed MFT keeps the asymptotic
significance level also in empirical point processes with a finite time
horizon, that is, rejecting the null hypothesis of stationary rate with
probability about $\alpha$. In contrast, the proposed MFA for change
point detection is a heuristic procedure that is not associated with a
specific significance level. However, as mentioned in Section~\ref
{sectSWD}, the SFA is a common method which yields consistent change
point estimates under mild conditions in Gaussian models [\citet
{Huskova2001}]. In addition, we explain here why the MFA, after taking
into account the typical number of falsely detected change points
(false positives, FP), should not overestimate the number of true
change points.
More precisely, a true change point does not increase the number of
FPs. This is because a true change point can only affect its
$h$-neighborhood, which is cut out in the SFA after detection. Outside
this $h$-neighborhood, the remaining process should resemble a process
derived under the null hypothesis, and thus produce about as many FPs
as under the null hypothesis with the same threshold $Q$. For the MFA
with multiple windows, a similar argument holds because change points
are only added when no accepted change point lies within their
$h$-neighborhood (cf.~Section~\ref{sectMFA}). Thus, one change point
should usually lead to at most one detection.

In order to support these considerations, Table~\ref{tableFDR} shows
simulation results of Gamma processes of length $T=700$ with a change
point at $c=350$ in which we investigate the number of correctly and of
falsely detected change points. A~change point is called correctly
detected if its $h$-neighborhood overlaps a true change point, whereas
$h$ corresponds to the window used for detection in the MFA. Rate
changes of different heights are simulated in order to account for
different detection probabilities of the inserted change point (first
column). In this setting, the MFA does not falsely detect more change
points than under the null hypothesis. The number of FPs (second
column) and the number of processes with at least one falsely detected
change point (third column) even decrease slightly because after
cutting $h$-neighborhoods, the remaining process is shorter, and thus
less likely to cross the threshold by chance.

%
\begin{table}
\tabcolsep=0pt
\caption{Simulation results of $\Gamma(2,\lambda)$-processes of length
$T=700$ with a rate change at $c=350$. Life times are $\Gamma
(2,24)$-distributed on $[0,350)$ and $\Gamma(2,\lambda_i)$-distributed
on $[350,700]$ with $\lambda_i\in\{25,26,28,30\}$. The respective rates
are given on the left, 10,000 simulations per scenario}
\label{tableFDR}
\begin{tabular*}{\tablewidth}{@{\extracolsep{\fill}}@{}lcccc@{}}
\hline
\textbf{Rates} &  \textbf{Detection prob.} & \textbf{Mean number of FPs} & \textbf{\% of processes}\\
&  \textbf{of true cp} & \textbf{per process} & \textbf{with $\bolds{\ge}$1 FP} \\
\hline
$12 \leadsto12.5$ &0.119 &0.051 & 4.9\\
$12 \leadsto13$ &0.653 &0.048 & 4.6 \\
$12 \leadsto14$ &0.996 &0.050 & 4.9 \\
$12 \leadsto15$ &0.999 &0.048 & 4.6 \\
\hline
\end{tabular*}
\end{table}

\subsection{Multiple filters increase the detection probability}\label{mfitp}
We have already seen in the example in Section~\ref{subsecapplicatonofcpd} that multiple windows can increase the
probability to detect a change point. One explanation is that the
simultaneous use of multiple filters avoids the problem of choosing the
most appropriate single window size. But, more importantly, the
combination of multiple filters is advantageous because large windows
have a higher detection probability, whereas small windows can be more
precise or sensitive to fast changes. Accordingly, we show here in
simulations that the MFA can even detect more change points than the
best single window.

%
\begin{figure}

\includegraphics{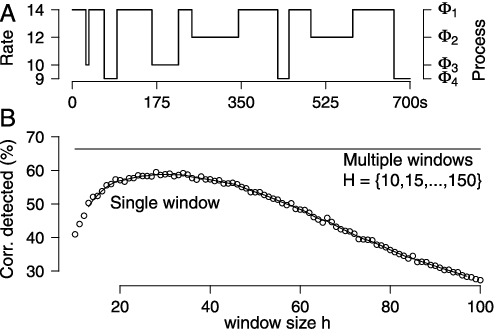}

\caption{Multiple filtering increases detection rate. \textup{(A)} A random
realization of the rate of a process $\Phi$ used in the simulations.
Intervals between change points are independent and
unif$(0,100]$-distributed. Simulated processes $\Phi_1,\ldots,\Phi_4$
have independent and $\Gamma(2,\lambda_i)$-distributed life times with
rate parameters $\lambda_1=28$, $\lambda_2=24$, $\lambda_3=20$ and
$\lambda_4=18$, leading to rates $\mu_1^{-1}=14$, $\mu_2^{-1}=12$, $\mu
_3^{-1}=10$ and $\mu_4^{-1}=9$ (indicated on the left). \textup{(B)} Mean
relative frequency of correct change point detections in simulated
processes as a function of the chosen window size. The points represent
the mean percentages of correct detections derived in $1000$
simulations using SFA. The curve shows a filtered average. The window
size with maximal detection probability of about $0.59$ is about
$\tilde h \approx28$. The horizontal line marks the mean relative
detection probability of about $0.66$ for a set of multiple windows
$H=\{10,15,\ldots,150\}$.}\label{improvementmultiple}
\end{figure}

In order to quantify this effect, we investigate the following random
change point model. We simulate processes $\Phi$ on $(0,700]$ in which
the rate fluctuates between four different values. The model includes
rate changes of different size and in different time scales. Each
process $\Phi$ is a piecewise composition of four independent renewal
processes $\Phi_1,\ldots,\Phi_4$ with Gamma-distributed life times with
event rates $\mu_1^{-1}=14$, $\mu_2^{-1}=12$, $\mu_3^{-1}=10$ and $\mu
_4^{-1}=9$. The change points for switches between the processes $\Phi
_1$ to $\Phi_4$ are given by a stationary renewal process $\Phi_c$ on
$(0,T]$ with change points $c_1,\ldots,c_{\llvert  \Phi_c\rrvert  }$. In order to
simulate change points in different time scales, the life times of $\Phi
_c$ are uniformly distributed on $[0,100]$. The observed process $\Phi$
is constructed from $\Phi_1,\ldots,\Phi_4$ as follows: set $\Phi\mid
\mid _{(0,c_1]}:=\Phi_{1}\mid_{(0,c_1]}$, that is, start in process $\Phi
_1$. At the first change point $c_1$ choose independently and uniformly
a process from $\{\Phi_2,\Phi_3,\Phi_4\}$ and jump into this process,
such that, for example, $\Phi\mid_{(c_1,c_2]}=\Phi_{2}\mid
_{(c_1,c_2]}$. Third, jump back deterministically to $\Phi_1$ at $c_2$,
that is, set $\Phi\mid_{(c_2,c_3]}=\Phi_{1}\mid_{(c_2,c_3]}$.
Repeat the procedure, choosing uniformly a process from $\{\Phi_2,\Phi
_3,\Phi_4\}$ at odd-valued change points and returning to $\Phi_1$ at
even-valued change points. An example of the rate of the resulting
process $\Phi$ is shown in Figure~\ref{improvementmultiple}A.

Figure~\ref{improvementmultiple}B indicates the percentage of
correctly detected change points in $1000$ simulations of the described
processes. A change point is called correctly detected if its
$h$-neighborhood overlaps a true change point, whereas $h$ corresponds
to the window used for detection in the MFA. In order to identify the
best individual window, the detection rate for the SFA is shown as a
function of the window size $h\in\{10,11,\ldots,100\}$. The percentage
of correct detections is maximal at about $59\%$ for a window size of
about $\tilde h=28$. Using the MFA with a set of multiple windows
chosen here arbitrarily as $H=\{10, 15,\ldots,150\}$, the correct
detection rate increases to about $66\%$.

\section{Application to spike train recordings}\label{sectdataanalysis}
\subsection{Data analysis}
In this section we apply the proposed MFT to a data set of 72 empirical
spike train recordings that were reported partly in \citet
{Schiemann2012}. The recording time $T$ was 540--900 seconds per spike
train, and the mean firing rate was about 6 spikes per second. The
significance level was set to $\alpha=5\%$.

In order to choose the set of windows, we use the results from
Section~\ref{sectsignlevel}, Figure~\ref{rejprobgamma}. Briefly, a
mean number of about 100--200 events in the smallest window is
required in order to keep the asymptotic significance level for point
processes with medium irregularity. Therefore, we choose a minimal
window of $h_1=25$ for a mean rate of $6$ Hz and $H=\{
25,50,75,100,125,150\}$.

%
\begin{figure}

\includegraphics{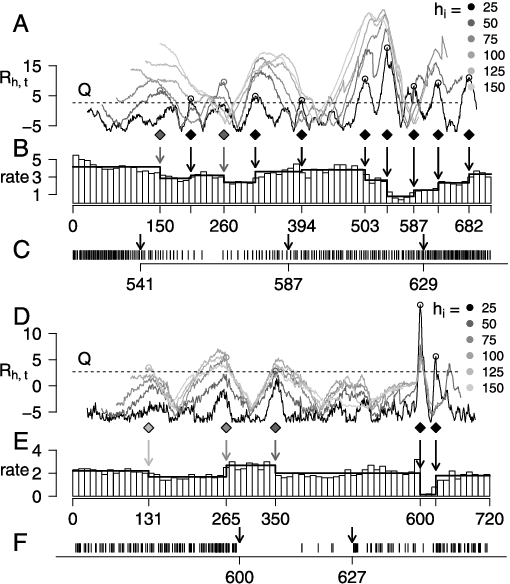}

\caption{Application of the MFT to two spike train recordings; $T=720$,
$H=\{25,50,75, 100,125,150\}, \alpha=5\%$. \textup{(A)} and \textup{(D)} The scaled
processes $(R_{h,t})_t$. Grayscales of the different window sizes are
indicated on the right. The dashed line marks the threshold $Q$, the
detected change points are marked by diamonds. In spike train $1$, $10$
change points are detected with two different windows; in spike train
$2$, $5$ change points are detected with four different windows. \textup{(B)}
and \textup{(E)} Rate histograms of the spike trains. Black step function
indicates estimated firing rate. \textup{(C)} and \textup{(F)} Short sections of the
spike trains. Arrows mark the estimated change points.}\label{application}
\end{figure}

\begin{figure}

\includegraphics{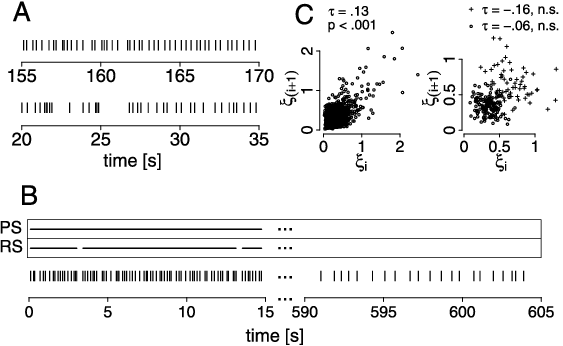}

\caption{Examples of data analyses applying the MFT. \textup{(A)} Sections of
spike train 1 (above) and 2 (below) indicating changes in the variance
of the life times. Left parts of illustrated sections show higher
irregularity than right parts. \textup{(B)} Application of Poisson Surprise (PS)
and Rank Surprise (RS) algorithms to spike train 1, using standard
parameters (only sections with surprise value $S>10$ marked as bursts
in both methods). Horizontal lines indicate bursts identified with PS
and RS methods, respectively. Because sections with high rate (left)
and low rate (right) are jointly analyzed, high-rate sections are
identified as long bursts. \textup{(C)} Analyzing serial correlations globally
yields spurious positive rank correlation (left), whereas serial
correlations within different sections (right, two sections indicated
by point characters) can be slightly negative and nonsignificant.}
\label{figdataana}
\end{figure}

Figure~\ref{application} shows two spike train analyses in which
multiple change points have been detected. As indicated by the
different grayscales, different window sizes were used for change point
estimation. From the set of $72$ spike trains, $62$ were identified as
nonstationary. In $50$ spike trains, at least two change points were
detected, and in $37$ spike trains, more than one window was necessary
for the detection of these change points.
Across all spike trains, the mean rate of detected change points was
about $0.32$ per minute. The lengths of intervals between detected
change points followed a right-skewed distribution with median $75$~s
and quartiles $q_1=44$~s and $q_3=123$~s. The height of a detected rate
change, measured as the difference of estimated rates $\hat\mu^{-1}_1$
and $\hat\mu^{-1}_2$ at the change point in relation to their mean,
$\llvert  \hat\mu^{-1}_1-\hat\mu^{-1}_2\rrvert  /(0.5 (\hat\mu^{-1}_1+\hat\mu
^{-1}_2))$, ranged between about $0.5$\% and $173$\%. As one can see
from the illustrations of the rate profiles in Figure~\ref{application}B~and~E,
the estimated rate profile corresponds well to a
rate estimate that is obtained from visual inspection. Figure~\ref{figdataana}A illustrates that both spike trains show varying
variance in their inter-spike intervals.

The identification of changes in the firing rate within neuronal spike
trains can facilitate their interpretation and avoid pitfalls. Most
importantly, the detected change points can be used for the separation
of a spike train into sections of virtually stationary firing rate.
This is important for multiple analysis techniques that assume rate
stationarity for the description and statistical analysis of single or
multiple spike trains, for example, techniques that study temporal
coordination between processes [e.g., \citet{Gruen02b,Staude2008,Schneider08}]. Here, we show two simple
analysis examples for individual spike trains.

%

First, variability of variance in the inter-spike intervals in dopamine
(DA) neurons is often expressed as a switching of firing between a
low-rate single spike background pattern and short events with
relatively many spikes, so-called ``bursts'' (cf.~also Figure~\ref{figdataana}A, bottom spike train: higher irregularity in the left
part). For DA neurons, burst firing has been shown to possess important
behavioral significance, as it is coupled to an increase of DA release
[\citet{Gonon1988,Redgrave2010,Schiemann2012}]. Such bursts
usually span very short periods with up to about $10$ spikes and can
thus not be detected with the asymptotic MFT, which requires about
100--200 spikes per window. However, the MFT can be an essential
preprocessing step in burst detection when existing methods require
rate stationarity. In two common methods for burst detection, bursts
are described as short periods with ``surprisingly many'' spikes
[\citet{Legendy1985,Gourevitch2007}]. These methods, called
Poisson Surprise (PS) and Rank Surprise (RS), assume rate stationary
Poisson or renewal processes and identify the ``surprising'' nature of
a burst by comparison to the overall mean life time. If periods of
different rates are jointly analyzed, the number of spikes in high-rate
sections that are assigned to bursts can be much larger than when
applying the algorithms to separate sections with approximately
stationary firing rate. Figure~\ref{figdataana}B illustrates this
effect by exemplary application of PS and RS burst detection algorithms
to spike train 1, for which visual inspection indicates nonbursty
firing activity (see also Figures~\ref{application}C, \ref{figdataana}A, top panel).
The horizontal lines in Figure~\ref{figdataana}B indicated by PS and RS indicate the bursts identified
by applying the two methods to the whole spike train. Almost all spikes
in the high-rate section are assigned to long ``bursts'' consisting of
$50$ and more spikes (in illustrated sections: PS: one long burst with
$160$ spikes, RS: three long bursts with 18, 54 and 38 spikes). This
is, however, inconsistent with the assumed physiological function and
short duration of DA bursts. In agreement with these considerations,
practically no bursts are identified in spike train 1 when applying the
MFT first and separately analyzing the sections with different rates
(PS: no bursts, RS: one burst with three spikes, not in illustrated
section). Thus, by separating between multiple longer sections of
different and unknown firing rates, the present algorithm for change
point detection complements burst detection methods which aim at
separating the two states ``bursty'' and ``nonbursty'' [e.g.,
\citet{Tokdar2010}].

Second, rate changes might cause potential misinterpretations of serial
correlations of life times, which has also been discussed in the
context of neuronal spike train analysis by \citet{Farkhooi2009}.
Consider a renewal process consisting of two periods with different
rates. In each period, correlation between adjacent life times is zero,
but in the high-rate section, short life times follow short life times,
and in the low-rate section, long life times follow long life times.
This induces a positive correlation in the global analysis. A similar
result is obtained in spike train 1, in which we exemplarily analyze
the correlation between adjacent life times with Kendall's rank
correlation $\tau$ (Figure~\ref{figdataana}C). A global analysis
falsely indicates a significant positive correlation (left panel, $\tau
=0.13$, $p<0.001$) due to rate changes, whereas most correlations in
individual sections are slightly negative and not significantly
different from zero. The right part of Figure~\ref{figdataana}C shows
two separate data pieces with different rates and slightly negative
correlations and illustrates how the joint analysis of such data sets
can produce a spurious positive global correlation. Because serial
correlations may reflect intrinsic neuronal properties [\citet
{Benda2003}], the application of the MFT as a preprocessing step can
also be helpful in this context.

Finally, apart from improving statistical analysis by detecting periods
of roughly constant rate, the detected rate changes themselves might
contain important information. For example, in addition to bursts,
periods of very low rate (``pauses,'' see, e.g.,~Figure~\ref{application}F) may also have behavioral relevance. A recent study
showed that the duration of these periods in DA neurons can be
associated with the expression of fear [\citet{Mileykovskiy2011}],
and a modeling study demonstrated that synchronized pauses in spiking
activity of many DA neurons can reduce information transmission in DA
type 2 receptors [\citet{Dreyer2010}]. In addition to pauses, more
complex change point sequences, such as multiple successive increases
in the firing rate, could reflect specific prolonged changes in the
typical DA activity that have been described recently [\citet{Howe2013}].

\subsection{Practical issues and R-code}
In practice, the described procedures can be applied easily. Depending
on a rough estimate of the overall rate and irregularity of the
process, one needs to choose the smallest window such that the
asymptotic properties are kept. One can then choose a set of windows up
to the largest interesting time scale. Then, the threshold $Q$ can be
estimated by repeated simulation of the limit process $(L_{h,t})_t$
[equation (\ref{supremelimit})].

In the supplementary material \citet{MesserCode} we provide an R
code that performs these steps efficiently within one single routine
and returns an illustration comparable to Figure~\ref{application}. It
also suggests a set of window sizes for a given time series of events.
The code can be applied easily, using as input only a time series of
events and (optional) a significance level and a set of windows, and
returning a set of estimated change points.

\section{Discussion}
In this paper we have developed a multiple filter technique for the
detection of change points in the event rate of time series. Motivated
by the problem that rate stationarity of the underlying processes is
crucial to many statistical analysis techniques, the multiple filter
test (MFT) tests the null hypothesis of rate stationarity against the
alternative of finitely many change points. In a second step, a
multiple filter algorithm (MFA) identifies and locates an unspecified
number of change points in the rate of the process. In addition, it
includes a graphical representation in which strong deviations from
rate stationarity can be visualized.

As a first extension to existent approaches, we introduce a general
class of point processes called renewal processes with varying variance
(RPVV). In addition to standard renewal assumptions reflected, for
example, in Poisson or Gamma processes, an RPVV assumes that the
variance of life times can show a certain degree of variability, which
includes, for example, mixtures of Gamma processes in the simplest
case. We propose RPVVs in order to account for the high variability of
patterns observed empirically and to allow for the detection of rate
changes irrespective of variance changes, which may be analyzed in
subsequent, separate steps when rate changes have been identified.

In order to test the null hypothesis of rate stationarity against the
alternative of finitely many change points, we extend a standard
filtered derivative method which compares the number of events in
adjacent windows in a moving window manner. Due to the general RPVV
assumptions, statistical significance of deviations from rate
stationarity cannot be tested by standard bootstrap approaches because
the life times are not necessarily identically distributed. Therefore,
we extend an asymptotic result of \citet{Steinebach1995} to RPVVs
and show that the limit $(L_{h,t})_t$ of the filtered derivative
process is a $2h$-dependent and zero mean Gaussian process. Notably,
this limit $(L_{h,t})_t$ is independent of the underlying RPVV
parameters such as the rate or the variances. By using the limit
process, thresholds for testing the statistical significance of
deviations from rate stationarity can be obtained by simulation.

As a second extension to existent approaches, we combine multiple
window sizes $h\in H$ in order to detect rate changes at fast and slow
time scales simultaneously. In the present asymptotic setting, multiple
window sizes can be combined easily because the set of processes $\{
(G_{h,t})_t\mid  h\in H\}$ depends on one underlying RPVV. In the same way,
the set of limit processes $\{(L_{h,t})_t \mid  h\in H\}$ depends on one
underlying Brownian motion. In addition, the use of multiple windows
requires two considerations: first, the statistical properties of
$(L_{h,t})_t$ depend on the window size $h$. Therefore, we standardize
the processes in order to give similar impact to every window size $h$.
Second, change point detection requires an extended algorithm that
combines the change points detected by multiple windows. Our multiple
filter algorithm is based on the idea of preferring change points
estimated by smaller windows to those estimated by larger windows. In a
random change point model with multiple time scales that we used here
for simulation, the MFA could detect more rate changes than the best
individual window.

The presented methods can be particularly relevant for practical
applications. First, the general assumptions of RPVVs cover a high
variability of patterns observed in empirical time series. Second,
multiple filtering can take into account that rate changes in empirical
time series can occur at fast and slow time scales simultaneously.
In practice, one should keep in mind that the MFA always estimates a
step function even when applied to a rate profile with gradual changes,
and that very short time scales, for example, bursts with a few spikes,
cannot be investigated by this asymptotic method. Third, in order to
enable an easy application of the MFA, we provide an R code that
includes all necessary steps within one single routine. It can be
computed efficiently, and it also includes a graphical illustration of
the resulting filtered derivative processes, in which large values
indicate deviations from rate stationarity. In an exemplary application
of the MFA to single unit neuronal recordings,
we illustrate that the detection of rate changes can be important for
the understanding of neuronal information processing and show that the
MFA can be a useful preprocessing step for data analysis techniques
that assume rate stationarity.

In summary, we believe that the present multiple filter technique can
be useful for the estimation of change points in the event rate of time
series of events. It may be used as a universal preprocessing step
whenever statistical analysis methods are sensitive to deviations from
rate stationarity.

%
\begin{appendix}\label{app}
\section*{Appendix}

In this \hyperref[app]{Appendix} we prove Theorem~\ref{hauptaussagegeneralisiert}.
Main ingredients of this proof are first the convergence of the
normalized counting process $(N_t)_{t\ge0}$ which is shown in
Section~\ref{convresccount} (cf.~Proposition~\ref{proprescaledcount}), and second the consistency of the estimator
$(\hat{s}^2)_{t\in\tau_h}$ defined in (\ref{schaetzers}). This is
shown in Section~\ref{convesti} (cf.~Proposition~\ref{propconvergenceshat}). First, in Section~\ref
{technicalpreliminaries} elementary facts are collected which are
later needed repeatedly and for which we do not claim originality. The
pieces are finally put together in Section~\ref{proofhauptaussagegeneralisiert} to prove Theorem~\ref
{hauptaussagegeneralisiert}.

The following notation is used:
for $\tau>0$ the set of all real-valued continuous functions on
$[0,\tau]$ is denoted by $C[0,\tau]$ and the set of all c\`adl\`ag
functions by $D[0,\tau]$. We abbreviate the metric induced by the
supremum norm by $d_{\llVert  \cdot\rrVert  }$, the Skorokhod metric on $D[0,\tau]$
by $d_\mathrm{SK}$. Analogously, we define $C[0,\infty)$ and use the
metric $d_{\llVert  \cdot\rrVert  }$ which induces the topology of compact
convergence. Further, we use $D[0,\infty)$ and $D[h,T-h]$ with
$d_{\mathrm{SK}}$. Note that convergence in $(D[0,\infty),d_{ \llVert  \cdot\rrVert  })$
implies convergence in $(D[0,\infty),d_\mathrm{SK})$.

\subsection{Technical preliminaries}\label{technicalpreliminaries}
The lemmas in this subsection have different assumptions on the renewal
processes occurring. However, note that the assumptions of all lemmas
in this subsection are fulfilled for an RPVV as in Definition~\ref{defgrp}.

First, we want to assure that the number of events $N_t$ in an RPVV
tends to infinity almost surely (a.s.), while explosion is avoided.

%
\begin{lemm}\label{lemm3}
Let $\{\xi_i\}_{i\ge1}$ be a sequence of independent, positive,
integrable random variables, interpreted as the life times of a point
process on the positive line, and $(N_t)_{t\ge0}$ the associated
counting process as in (\ref{defnt}). Then we have almost surely
%
\begin{equation}
\label{ntinf} N_t \to\infty\qquad(t\to\infty).
\end{equation}
If the $\{\xi_i\}_{i\ge1}$ are square integrable and satisfying
conditions (\ref{nullconda}) and (\ref{nullcondc}), then for all
$t\ge0$ we have almost surely
%
\begin{equation}
\label{ntfin} N_t < \infty.
\end{equation}
\end{lemm}

\begin{pf}
For (\ref{ntinf}) note that $N_t$ is increasing in $t$. For all fixed
$k>0$ we have
\[
\{ N_n < k \} = \Biggl\{\sum_{i=1}^k
\xi_i > n \Biggr\} \downarrow \bigcap_{n\ge1}
\Biggl\{\sum_{i=1}^k \xi_i > n
\Biggr\} \qquad(n\to \infty).
\]
Since the $\xi_i$ are integrable, we have $P(\xi_i<\infty)=1$.
Continuity from above (applied twice) yields
\begin{eqnarray*}
P \biggl(\bigcap_{n\in\mathbb{N}}\{N_n<k\} \biggr)
&=& \lim_{n\to\infty} P \Biggl(\sum_{i=1}^k
\xi_i > n \Biggr)
\le\lim_{n\to\infty} P \Biggl(\bigcup
_{i=1}^{k} \biggl\{\xi_i >
\frac{n}{k} \biggr\} \Biggr)
\\
&\le&\lim_{n\to\infty} \sum_{i=1}^{k}
P \biggl(\xi_i > \frac{n}{k} \biggr) = 0.
\end{eqnarray*}
This implies (\ref{ntinf}).

For (\ref{ntfin}) first note that (\ref{nullconda}) and (\ref
{nullcondc}) imply Kolmogorov's conditions (\ref{kol1})--(\ref{kol3})
with $\xi_i^2$ replaced by $\xi_i$. Hence, we have the SLLN for $\{\xi
_i\}_{i\ge1}$, that is, $(1/n)\sum_{i=1}^n \xi_i \to\mu$ a.s. as $n\to
\infty$. This implies $\sum_{i=1}^n \xi_i \to\infty$ a.s. and
\[
P(N_t<\infty) = P \Biggl(\bigcup_{n=1}^\infty
\Biggl\{\sum_{i=1}^n \xi_i > t
\Biggr\} \Biggr) = P \Biggl(\lim_{n\to\infty} \sum
_{i=1}^n \xi_i > t \Biggr) =1.
\]\upqed
\end{pf}

Now we show that the number of events in successively increased
windows, scaled with the widths of the windows, tends to the stationary
rate $1/\mu$ almost surely.

%
\begin{lemm}\label{lemm4}
Let $\{\xi_i\}_{i\ge1}$ be a sequence of independent, positive,
square-integrable random variables satisfying conditions (\ref{nullconda}) and (\ref{nullcondc}), which are interpreted as the
life times of a point process on the positive line, and $(N_t)_{t\ge
0}$ the associated counting process as in (\ref{defnt}).
Then for all $0\le s< t$ we have, as $n\to\infty$, almost surely
\[
\frac{N_{nt} - N_{ns}}{n(t-s)} \longrightarrow\frac{1}{\mu}.
\]
\end{lemm}

\begin{pf} As in the proof of Lemma~\ref{lemm3}, conditions (\ref{nullconda}) and (\ref{nullcondc}) imply the SLLN for $\{\xi_i\}
_{i\ge1}$, that is, with $S_n = \sum_{i=1}^n\xi_i$ for $n\ge1$, we have
$S_n/ n \to\mu$ a.s. for $n\to\infty$. By Lemma~\ref{lemm3} we have
$N_t \to\infty$ a.s.~as $t\to\infty$, hence, $S_{N_t}/N_t\to\mu$
a.s.~as $t\to\infty$. Now, for all $t\ge0$ we find $S_{N_t} \le t \le
S_{N_t + 1}$, so that (for all $t$ sufficiently large such that $N_t\ge1$)
\[
\frac{S_{N_t}}{N_t} \le\frac{t}{N_t} \le\frac{S_{N_t+1}}{N_t+1}
\frac
{N_t+1}{N_t}.
\]
Since the left-hand side and the right-hand side tend to $\mu$ a.s.,~we
obtain $N_t / t \to1/\mu$ a.s.~as $t\to\infty$. This implies, as $n\to
\infty$, almost surely
\[
\frac{N_{nt} - N_{ns}}{n(t-s)} = \frac{t}{t-s}\frac{N_{nt}}{nt} - \frac
{s}{t-s}
\frac{N_{ns}}{ns} \longrightarrow\frac{t}{t-s}\frac{1}{\mu} -
\frac{s}{t-s}\frac{1}{\mu} = \frac{1}{\mu}.
\]\upqed
\end{pf}


The next result will secure that the events in the different windows
will evolve properly in time.

%
\begin{lemm}\label{lemm5}
Let $(N_t)_{t \ge0}$ be a counting process with $N_0=0$ such that for
some $\mu>0$ and for all $0\le s < t$ we have $N_{nt}-N_{ns} \sim
n(t-s)/\mu$ almost surely. Further, let $V_1,V_2,\ldots$ be a sequence
of independent random variables that satisfies the SLLN. Then for all
$0\le s < t$ we have, as $n\to\infty$, almost surely
\[
\frac{1}{N_{nt}-N_{ns}}\sum_{i=N_{ns}+1}^{N_{nt}}
V_i \longrightarrow c.
\]
\end{lemm}

\begin{pf}
Note that choosing $s=0$ in the statement of the lemma
implies $N_t\sim t/\mu$ a.s., such that we find $N_t\to\infty$ as
$t\to\infty$. Then we calculate (for $N_{ns}>0$, the case $N_{ns}=0$
being similar)
\[
\frac{1}{N_{nt}}\sum_{i=1}^{N_{nt}}
V_i = \frac{N_{ns}}{N_{nt}} \frac
{1}{N_{ns}} \sum
_{i=1}^{N_{ns}} V_i + \frac{N_{nt}-N_{ns}}{N_{nt}}
\frac
{1}{N_{nt}-N_{ns}} \sum_{i=N_{ns}+1}^{N_{nt}}
V_i,
\]
so that, for $n\to\infty$,
\begin{eqnarray*}
&& \frac{1}{N_{nt}-N_{ns}} \sum_{i=N_{ns}+1}^{N_{nt}}
V_i \\
&&\qquad  = \frac
{N_{nt}}{N_{nt}-N_{ns}} \Biggl( \frac{1}{N_{nt}}\sum
_{i=1}^{N_{nt}} V_i - \frac{N_{ns}}{N_{nt}}
\frac{1}{N_{ns}} \sum_{i=1}^{N_{ns}}
V_i \Biggr)
\\
&&\qquad \longrightarrow \frac{t}{t-s} \biggl(c - \frac{s}{t} c \biggr) = c
\qquad \mbox{a.s.}
\end{eqnarray*}\upqed
\end{pf}

%
%
\begin{coro}\label{coro1}
Let $(v_i)_{i\ge1}$ be a sequence in $\mathbb{R}$ with $(1/n)\sum_{i=1}^n v_i \to c$ as $n\to\infty$. Then for all $ 0\le s<t $, as $n\to
\infty$, we have
\[
\frac{1}{n(t-s)}\sum_{i=\lfloor ns \rfloor+1}^{\lfloor nt \rfloor}
v_i \to c.
\]
\end{coro}

Finally, we provide a result related to Lemma~\ref{lemm5} for the
Lindeberg condition which will be used below to apply the
Lindeberg--Feller CLT for triangular schemes.

\begin{lemm}\label{lemm2}
Let $\{\xi_i\}_{i\ge1}$ be a sequence of independent, square-integrable
random variables satisfying conditions (\ref{nullconda}), (\ref
{nullcondd}) and (\ref{nullcondb}). Then, for all $0\le s< t$ and
all $\varepsilon>0$ we have, as $n\to\infty$,
\[
\frac{1}{s_n^2(s,t)} \sum_{i=\lfloor ns \rfloor+1}^{\lfloor nt \rfloor
}
\mathbb{E} \bigl[(\xi_i-\mu)^2 \mathbbm{1}_{\{(\xi_i-\mu)^2>\varepsilon
^2 s_n^2(s,t)\}}
\bigr] \longrightarrow0,
\]
where
$s_n^2(s,t):= \sum_{i=\lfloor ns \rfloor+1}^{\lfloor nt \rfloor} \Var(\xi_i)$.
\end{lemm}

\begin{pf}
Let $0\le s<t$. Condition (\ref{nullcondd}) and Corollary~\ref{coro1}
imply, as $n\to\infty$,
\[
s_n^2(s,t)\sim(t-s)\sigma^2 n \sim
\frac{t-s}{t} \sum_{i=1}^{\lfloor
nt \rfloor} \Var(\xi_i).
\]
For $\varepsilon>0$ set $\eta:= \varepsilon\sqrt{(t-s)/(2t)}$. It
follows the existence of an element $n_0 = n_0(s,t) \in\mathbb{N}$ so
that for all $n> n_0$ and an appropriate null sequence $o(1)$ we have
\begin{eqnarray*}
\varepsilon^2 s_n^2(s,t) & =& \bigl(1+o(1)
\bigr)\varepsilon^2 \frac{t-s}{t} \sum
_{i=1}^{\lfloor nt \rfloor}\Var(\xi_i) >
\eta^2 \sum_{i=1}^{\lfloor nt \rfloor}\Var(\xi_i).
\end{eqnarray*}
Thus, for $n> n_0$ we obtain
\begin{eqnarray*}
&& \frac{1}{s_n^2(s,t)} \sum_{i=\lfloor ns \rfloor+1}^{\lfloor nt \rfloor
}\mathbb{E} \bigl[(\xi_i-\mu)^2 \mathbbm{1}_{\{(\xi_i-\mu)^2>\varepsilon
^2 s_n^2(s,t)\}}
\bigr]
\\
&&\qquad \le\bigl(1+o(1)\bigr)\frac{t}{t-s} \frac{1}{\sum_{i=1}^{\lfloor nt \rfloor
}\Var(\xi_i)}\sum
_{i=1}^{\lfloor nt \rfloor}\mathbb{E} \bigl[(\xi_i-
\mu)^2 \mathbbm{1}_{\{(\xi_i-\mu)^2>\varepsilon^2 s_n^2(s,t)\}
} \bigr]
\\
&&\qquad \le\bigl(1+o(1)\bigr)\frac{t}{t-s} \frac{1}{\sum_{i=1}^{\lfloor nt \rfloor
}\Var(\xi_i)}\sum
_{i=1}^{\lfloor nt \rfloor}\mathbb{E} \bigl[(\xi_i-
\mu)^2 \mathbbm{1}_{\{(\xi_i-\mu)^2>\eta^2\sum_{i=1}^{\lfloor
nt \rfloor}\Var(\xi_i) \}} \bigr],
\end{eqnarray*}
and the last expression tends to zero due to condition (\ref{nullcondb}).
\end{pf}

\subsection{Convergence of the rescaled counting process}\label{convresccount}
In this subsection we show that the counting process $(N_{t})_{t\ge0}$
as in (\ref{defnt}) properly normalized converges weakly to a standard
Brownian motion.

For an RPVV $\Phi$ with parameters $\mu$ and $\sigma^2$, the rescaled
version of the corresponding counting process $(N_t)_{t\ge0}$ is given by
%
\begin{equation}
\label{rn261113} Z_t^{(n)}:= \frac{N_{n t} - \nicefrac{n t}{\mu}}{\sqrt{n}\sqrt
{\nicefrac{\sigma^2}{\mu^3}}},\qquad t
\ge0.
\end{equation}
The present subsection is devoted to the proof of this proposition:

\begin{prop}\label{proprescaledcount}
Let $\Phi$ be an RPVV with associated parameters $\mu$ and $\sigma^2$.
Further, let $(W_t)_{t\ge0}$ be a standard Brownian motion. Then, in
$(D[0,\infty), d_{\mathrm{SK}})$ we have the convergence, as $n\to\infty$, in
distribution
\[
\bigl(Z_t^{(n)}\bigr)_{t\ge0} \stackrel{d} {
\longrightarrow}(W_t)_{t\ge0}.
\]
\end{prop}

For the proof of Proposition~\ref{proprescaledcount} note that we
have the following result from \citet{Billingsley1999}, Theorem~14.6:

\begin{prop}\label{renewalprop}
Let $\{\xi_i\}_{i\ge1}$ be a sequence of positive random variables and
$(W_t)_{t\ge0}$ be a standard Brownian motion. Assume the existence of
positive constants $\mu$ and $\sigma$, so that the rescaled process
$(X_t^{(n)})_{t\ge0}$ defined via
%
\begin{equation}
\label{waitingtimecondition} X_t^{(n)}: = \frac{1}{\sigma\sqrt{n}} \sum
_{i=1}^{[n t]}(\xi_i - \mu ),\qquad
t\ge0,
\end{equation}
converges weakly to $(W_t)_{t\ge0}$ in $(D[0,\infty), d_{\mathrm{SK}})$. Then,
the rescaled counting process $Z^{(n)}:= (Z_t^{(n)})_{t\ge0}$ defined
in (\ref{rn261113}) converges weakly to $(W_t)_{t\ge0}$ in $(D[0,\infty
), d_{\mathrm{SK}})$.
\end{prop}


Since convergence in $(D[0,\infty), d_{\llVert  \cdot\rrVert  })$ implies
convergence in $(D[0,\infty), d_{\mathrm{SK}})$, Proposition~\ref{proprescaledcount} is proved if the conditions in Proposition~\ref
{renewalprop} are satisfied. Thus, it remains to show the following
proposition:

\begin{prop}\label{lemmconvxn}
Let $\Phi$ be an RPVV with associated parameters $\mu$ and $\sigma^2$
and corresponding life times $\{\xi_i\}_{i\ge1}$. For $n=1,2,\ldots$
let the processes $X^{(n)}$ be defined as in (\ref{waitingtimecondition}). Then it holds in $(D[0,\infty), d_{\llVert  \cdot\rrVert })$ as $n\to\infty$ that
\[
\bigl(X_t^{(n)}\bigr)_{t\ge0} \stackrel{d} {
\longrightarrow}(W_t)_{t\ge0}.
\]
\end{prop}

For the proof of Proposition~\ref{lemmconvxn} we first show that
$(X_t^{(n)})_{t\in[0,\tau]}$ converges weakly to $(W_t)_{t\in[0,\tau]}$
in $(D[0,\tau], d_{\llVert  \cdot\rrVert  })$ for $\tau>0$, which is the subject of
the following Lemma~\ref{lemmconvxnrestr}. Afterward, we present the
proof of Proposition~\ref{lemmconvxn}, which then merely consists of
extending the result of Lemma~\ref{lemmconvxnrestr} from the
interval $[0,\tau]$ to $[0,\infty)$.

%
\begin{lemm}\label{lemmconvxnrestr}
Let $\Phi$ be an RPVV with associated parameters $\mu$ and $\sigma^2$
and corresponding life times $\{\xi_i\}_{i\ge1}$ and $\tau>0$. For
$n=1,2,\ldots$ let the processes $(X_t^{(n)})_{t\ge0}$ be defined as
in (\ref{waitingtimecondition}). Then it holds in $(D[0,\tau],
d_{\llVert  \cdot\rrVert  })$ as $n\to\infty$ that
\[
\bigl(X_t^{(n)}\bigr)_{t\in[0,\tau]} \stackrel{d} {
\longrightarrow}(W_t)_{t\in
[0,\tau]}.
\]
\end{lemm}

For the proof of Lemma~\ref{lemmconvxnrestr} we use the following
construction of processes which connects $(X_t^{(n)})_{t\ge0}$ and its
restriction $(X_t^{(n)})_{t\in[0,\tau]}$ to the setting of RPVVs.

%
\begin{cons}\label{consyn}
Let $\Phi$ be an RPVV with corresponding parameters $\mu$ and $\sigma
^2$ and life times $\{\xi_i\}_{i\ge1}$. Let $(X_t^{(n)})_{t\ge0}$ be
constructed from $\{\xi_i\}_{i\ge1}$ as in~(\ref{waitingtimecondition}). For $n=1,2,\ldots$ and $\tau>0$ let the
restriction of time to $[0,\tau]$ be denoted by
$Y^{(n)}:=(X^{(n)})_{t\in[0,\tau]}$. Further denote the restriction of
the standard Brownian motion as $Y=(W_t)_{t\in[0,\tau]}$.
\end{cons}

To prepare the proof of Lemma~\ref{lemmconvxnrestr} note that we
have the following Theorem~\ref{theoprocessrestricted} from
\citet{Pollard1984}, Section~V, Theorem~19 (where we adjust the time
interval appropriately):

%
\begin{theo}\label{theoprocessrestricted}
Let $\tau>0$ and $Y,Y^{(1)},Y^{(2)},\ldots$ be random elements of
$(D[0,\tau], d_{\llVert  \cdot\rrVert  })$, each with independent life times.
Suppose $Y$ has continuous sample paths. Then, as $n\to\infty$, we have
$Y^{(n)} \stackrel{d}{\longrightarrow} Y$ in $(D[0,\tau],d_{\llVert  \cdot\rrVert
})$ if and only~if
\begin{longlist}[2.]
\item[1.] $Y_0^{(n)} \stackrel{d}{\longrightarrow} Y_0$.

\item[2.] For all $s,t$ with $0\le s<t\le\tau$ we have $Y_t^{(n)}-Y_s^{(n)}
\stackrel{d}{\longrightarrow} Y_t-Y_s$.

\item[3.] For all $\varepsilon>0$ there exist $\alpha>0,\beta>0$ and $n_0\in
\mathbb{N}$, such that $P(\llvert  Y_t^{(n)}-Y_s^{(n)}\rrvert  <\varepsilon)\ge\beta$
for all $t,s\in[0,\tau]$ with $0\le t-s <\alpha$ and all $n\ge n_0$.
\end{longlist}
\end{theo}

\begin{pf*}{Proof of Lemma~\ref{lemmconvxnrestr}}
We apply Theorem~\ref{theoprocessrestricted} to our setting of RPVVs:
the $Y,Y^{(1)},Y^{(2)},\ldots$ from Construction~\ref{consyn} have
independent increments and $Y$ has continuous sample paths. We now
verify that $Y,Y^{(1)},Y^{(2)},\ldots$ from Construction~\ref{consyn}
fulfill conditions 1--3 of Theorem~\ref{theoprocessrestricted}:
condition~1 is clear. For condition~2 note that for all $n\ge1$ and
all $0\le s<t\le\tau$ the increment $Y_t^{(n)}-Y_s^{(n)}$ is the sum of
elements of a triangular scheme. The $n$th row of this scheme is of the
type $\{(\xi_{i_{s_n}}-\mu)/\sigma\sqrt{n},
(\xi_{i_{s_n}+1}-\mu)/\sigma\sqrt{n},\ldots,
(\xi_{i_{t_n}}-\mu)/\sigma\sqrt{n}\}$, hence, it consists of
independent random variables.
For the variance of the increments we have
\begin{eqnarray*}
\Var\bigl(Y_t^{(n)}-Y_s^{(n)}
\bigr)
&=&\frac{1}{n\sigma^2} \sum_{i=\lfloor ns \rfloor+1}^{\lfloor nt \rfloor}
\Var(\xi_i)
\\
& =& (t-s) \frac{1}{\sigma^2}\frac{1}{n(t-s)}\sum_{i=\lfloor ns \rfloor
+1}^{\lfloor nt \rfloor}
\Var(\xi_i) \longrightarrow t-s,
\end{eqnarray*}
for $n\to\infty$,
where we use condition (\ref{nullcondd}) and Corollary~\ref{coro1}.

Due to condition (\ref{nullcondb}) and Lemma~\ref{lemm2}, the
Lindeberg condition is satisfied for the corresponding triangle scheme,
so that the Lindeberg--Feller CLT implies, as $n\to\infty$,
\[
Y_t^{(n)}-Y_s^{(n)} \stackrel{d} {
\longrightarrow} {\mathcal N}(0,t-s).
\]

Now, for condition 3 let $\varepsilon> 0$. For all $0\le s<t\le\tau$
Chebyshev's inequality implies
\begin{eqnarray*}
P\bigl(\bigl\llvert Y_t^{(n)}-Y_s^{(n)}
\bigr\rrvert <\varepsilon\bigr)
&=& 1 - P\bigl(\bigl\llvert Y_t^{(n)}-Y_s^{(n)}
\bigr\rrvert \ge \varepsilon\bigr)
\ge 1-\frac{1}{\varepsilon^2} \Var\bigl(Y_t^{(n)}-Y_s^{(n)}
\bigr)
\\
&=& 1 - \frac{1}{\varepsilon^2} \Biggl((t-s) \frac{1}{\sigma^2}\frac
{1}{n(t-s)}\sum
_{i=\lfloor ns \rfloor+1}^{\lfloor nt \rfloor}\Var(\xi_i)
\Biggr)
\\
&\ge& 1 - c_\varepsilon(t-s) =:\beta,
\end{eqnarray*}
where we use condition (\ref{nullcondc}), so that the constant
$c_\varepsilon$ does not depend on $s, t$ and $n$.
Now choose $\alpha>0$ sufficiently small such that $\beta>0$.

Hence, all conditions of Theorem~\ref{theoprocessrestricted} are
satisfied, thus, we obtain that the processes $Y^{(n)}$ converge weakly
to $Y=(W_t)_{t\in[0,\tau]}$ in $(D[0,\tau], d_{\llVert  \cdot\rrVert  })$ for $n\to
\infty$.
\end{pf*}


Finally, we extend Lemma~\ref{lemmconvxnrestr} to the time interval
$[0,\infty)$ and hence prove Proposition~\ref{lemmconvxn}. We use the
following theorem from \citet{Pollard1984}, Section~V, Theorem~23:

\begin{theo}\label{theoextendtoinfty}
Let $X,X^{(1)},X^{(2)},\ldots$ be random elements of $D[0,\infty)$,
with $X\in C$ a.s., for some separable set $C\subset(D[0,\infty),d_{\llVert  \cdot\rrVert  })$. Then, with convergence $n\to\infty$, the following
statements are equivalent:
%
\begin{eqnarray}
X^{(n)}&\stackrel{d} {\longrightarrow}& X\qquad\mbox{in } \bigl(D[0,\infty),d_{\llVert  \cdot\rrVert  }\bigr), \label{poll1}
\\
\bigl(X^{(n)}_t\bigr)_{t\in[0,\tau]} &\stackrel{d} {\longrightarrow}& (X_t)_{t\in
[0,\tau]}\qquad\mbox{in } \bigl(D[0,
\tau],d_{\llVert  \cdot\rrVert  }\bigr)\mbox{ for all }\tau>0. \label{poll2}
\end{eqnarray}
\end{theo}

\begin{pf*}{Proof of Proposition~\ref{lemmconvxn}}
We apply Theorem~\ref{theoextendtoinfty}:
let $X,X^{(n)},Y$ and $Y^{(n)}$ be derived from Construction~\ref{consyn}.
Note that $C[0,\infty)$ is a closed, separable subset of $(D[0,\infty
),d_{\llVert  \cdot\rrVert  })$; see \citet{Pollard1984}, page~108. Condition
(\ref{poll2}) has been shown in Lemma~\ref{lemmconvxnrestr}.
Hence, Theorem~\ref{theoextendtoinfty} applies and we obtain
Proposition~\ref{lemmconvxn}.
\end{pf*}

\subsection{Constistency of the estimators}\label{convesti}
Here we show the almost sure uniform convergence of our estimator
$(\hat s)_{t\in\tau_h}$ defined in equation (\ref{schaetzers}). This
will be needed for the proof of Theorem~\ref
{hauptaussagegeneralisiert} to exchange the denominator of
$G_{h,t}^{(n)}$ with\vspace*{1pt} an empirical normalization by application of
Slutsky's theorem. Note that for an a.s.~constant stochastic process in
$D[h,T-h]$, say, with constant $c$, we write $(c)_{t\in\tau_h}$.

We have the following consistency result for our estimator $(\hat s)_{t\in\tau_h}$:

%
\begin{prop}\label{propconvergenceshat}
Let $\Phi$ be an RPVV with corresponding parameters $\mu$ and $\sigma^2$.
Let $T>0$, $h\in(0,T/2]$ and $\hat s^2(t,h)$ be as defined in equation
(\ref{schaetzers}). Then we have in $(D[h,T-h], d_{\llVert  \cdot\rrVert  })$, as
$n\to\infty$, almost surely
\[
\biggl(\frac{\hat s^2(nt,nh)}{n} \biggr)_{t \in\tau_h } \longrightarrow \biggl(
\frac{2h\sigma^2}{\mu^3} \biggr)_{t \in\tau_h }.
\]
\end{prop}

\begin{pf}
We show the uniform a.s.~convergence\vspace*{1pt} of $(\hat{\mu
}_{\ler})_{t\in\tau_h}$ and $(\hat{\mu}_{\ri})_{t\in\tau_h}$ to the
constant $\mu$ in Lemma~\ref{lemmconvestimu}, and the uniform
a.s.~convergence of $(\hat{\sigma}^2_{\ler})_{t\in\tau_h}$ and $(\hat{\sigma}^2_{\ri})_{t\in\tau_h}$ to the constant $\sigma^2$ in Lemma~\ref
{lemmconvestisigma}. Uniform a.s.~convergence interchanges with sums
in general and with products if the limits are constant. Hence, Lemmas~\ref{lemmconvestimu} and~\ref{lemmconvestisigma} and the form
of the estimator $\hat s^2$ in (\ref{schaetzers}) imply the assertion.\vadjust{\goodbreak}
\end{pf}

In the rest of the section we show the uniform a.s.~convergence of the
estimators $(\hat\mu_{\ri})_{t\in\tau_h}$ and $(\hat\sigma_{\ri}^2)_{t\in
\tau_h}$, respectively, $(\hat\mu_{\ler})_{t\in\tau_h}$ and $(\hat\sigma
_{\ler}^2)_{t\in\tau_h}$ (see Lemmas~\ref{lemmconvestimu}~and~\ref{lemmconvestisigma}), as needed in the latter proof. We start with a
uniform a.s.~result for the scaled counting process $(N_t)_{t\ge0}$.

%
\begin{lemm}\label{lemmconvnt}
Let $\Phi$ be an RPVV with associated mean $\mu$.
Let $T>0$, $h\in(0,T/2]$. Then we have in $(D[h,T-h], d_{\llVert  \cdot\rrVert  })$
a.s.~as $n\to\infty$ that
%
\begin{eqnarray}
\label{convnt1} \biggl(\frac{N_{n(t+h)}-N_{nt}}{nh/\mu} \biggr)_{t\in\tau_h} &\longrightarrow& (1)_{t\in\tau_h},
\\
\label{convnt2} \biggl(\frac{N_{nt}-N_{n(t-h)}}{nh/\mu} \biggr)_{t\in\tau_h} &\longrightarrow& (1)_{t\in\tau_h}.
\end{eqnarray}
\end{lemm}

\begin{pf}
We show the first statement (\ref{convnt1}). The second one (\ref
{convnt2}) follows analogously.

We even prove that in $(D[0,T-h], d_{\llVert  \cdot\rrVert  })$ it holds a.s.~as $n\to
\infty$ that
%
\begin{equation}
\label{convnt3} \biggl(\frac{N_{n(t+h)}-N_{nt}}{nh/\mu} \biggr)_{t\in[0,T-h]}
\longrightarrow(1)_{t\in[0,T-h]}.
\end{equation}
It is sufficient to show that almost surely
%
\begin{eqnarray}\label{convnt4}
\lim_{n\to\infty} \sup_{t\in[0,T-h]}
\frac{N_{n(t+h)}-N_{nt}}{nh/\mu} &\le& 1\quad\mbox{and}
\nonumber\\[-8pt]\\[-8pt]\nonumber
\lim_{n\to\infty} \inf _{t\in[0,T-h]} \frac{N_{n(t+h)}-N_{nt}}{nh/\mu} &\ge&1.
\end{eqnarray}

In order to see the left inequality, we decompose the interval $(0,nT]$
into equidistant sections of length $n \varepsilon$. We use the notation
%
\begin{equation}
\label{rn271113} \bigl\llvert \lceil x \rceil\bigr\rrvert:=\lceil x \rceil+ 1,
\qquad \bigl\llvert \lfloor x \rfloor \bigr\rrvert:=\lfloor x \rfloor- 1, \qquad
x>0.
\end{equation}
Then each window $(nt,n(t+h)]$ for $t \in(h,T-h]$ is overlapped by one
of the finitely many windows $(kn\varepsilon,kn\varepsilon+n\llvert   \lceil
h/\varepsilon\rceil\rrvert  \varepsilon ]$ for $k=0,1,\ldots,\lceil
T/\varepsilon\rceil$. Therefore, we find for all $\varepsilon>0$
\[
\sup_{t\in[0,T-h]} N_{n(t+h)}-N_{nt} \le\max
_{k=0,1,\ldots,\lceil
T/\varepsilon\rceil} N_{kn\varepsilon+ n\llvert   \lceil h/\varepsilon\rceil
\rrvert  \varepsilon} - N_{kn\varepsilon}.
\]
Thus,
\begin{eqnarray*}
&& \sup_{t\in[0,T-h]} \frac{N_{n(t+h)}-N_{nt}}{nh/\mu}
\\
&&\qquad \le \max_{k=0,1,\ldots,\lceil T/\varepsilon\rceil} \frac{N_{kn\varepsilon
+ n\llvert   \lceil h/\varepsilon\rceil\rrvert  \varepsilon} - N_{kn\varepsilon
+nh}}{nh/\mu} + \max
_{k=0,1,\ldots,\lceil T/\varepsilon\rceil} \frac{N_{kn\varepsilon
+nh}-N_{kn\varepsilon}}{nh/\mu}.
\end{eqnarray*}
The first summand in the latter display becomes small, since $n\llvert   \lceil
h/\varepsilon\rceil\rrvert  \varepsilon\to nh$ for $\varepsilon\downarrow
0$. More precisely, for every $\delta> 0$, we can appropriately choose
$\varepsilon>0$, so that
\[
\max_{k=0,1,\ldots,\lceil T/\varepsilon\rceil} \frac{N_{kn\varepsilon
+ n\llvert   \lceil h/\varepsilon\rceil\rrvert  \varepsilon} - N_{kn\varepsilon
+nh}}{nh/\mu} \to\frac{\delta}{h}
\]
a.s.~as $n\to\infty$.
The second summand in the latter display converges to 1 a.s.~for $n\to
\infty$. This is because, due to Lemma~\ref{lemm4}, the convergence in
Lemma~\ref{lemmconvnt} is already known to hold a.s.~for finitely
many $t\in[0,T-h]$.
Thus, we find a.s.~that
\[
\lim_{n\to\infty} \sup_{t\in[0,T-h]} \frac{N_{n(t+h)}-N_{nt}}{nh/\mu}
\le\frac{\delta}{h} + 1.
\]
Since $\delta>0$ is arbitrary, for small $\varepsilon\downarrow0$ we
obtain a.s.~that
\[
\lim_{n\to\infty} \sup_{t\in[0,T-h]} \frac{N_{n(t+h)}-N_{nt}}{nh/\mu}
\le 1.
\]

For the right inequality of (\ref{convnt4}), we use the same
decomposition of the interval $(0,nT]$ into equidistant sections of
length $n \varepsilon$. Then each window $(nt,n(t+h)]$ for $t \in
(h,T-h]$ overlaps one of the finitely many windows $(kn\varepsilon,kn\varepsilon+ n\llvert   \lfloor h/\varepsilon\rfloor\rrvert  \varepsilon ]$ for
$k=0,1,\ldots,\lfloor(T-h)/\varepsilon\rfloor$. One can apply
arguments as for the proof of the left inequality of (\ref{convnt4})
to find that a.s. for $n\to\infty$
\[
\lim_{n\to\infty} \inf_{t\in[0,T-h]} (N_{n(t+h)}-N_{nt})/(nh/
\mu)\ge1.
\]
The assertion follows.
\end{pf}

Next we show the uniform a.s.~convergence of the estimators $(\hat\mu
_{\ri})_{t\in\tau_h}$, $(\hat\mu_{\ler})_{t\in\tau_h}$, $(\hat\sigma
_{\ri}^2)_{t\in\tau_h}$ and $(\hat\sigma_{\ler}^2)_{t\in\tau_h}$.
We use that uniform a.s.~convergence interchanges with sums in general
and with products if the limits are constant.
Recall the notation
\begin{eqnarray*}
\gamma &=& \gamma_{\ri}(nt,nh)
= \bigl\{\xi_i\dvtx
S_i,S_{i+1} \in\bigl(nt,n(t+h)\bigr], i=1,2,\ldots\bigr\}
\\
& =& \{\xi_i\dvtx  i = N_{nt}+2,\ldots, N_{n(t+h)}\}.
\end{eqnarray*}

We find our empirical quantities from equations (\ref{muschaetzer1})
and (\ref{sigmaschaetzer1}) as
%
\begin{eqnarray}
\hat\mu_{\ri} =\hat\mu_{\ri}(nt,nh)=\frac{1}{N_{n(t+h)}-N_{nt}-1}\sum
_{i=N_{nt}+2}^{N_{n(t+h)}}\xi_i
\nonumber\\[-8pt]\\[-8pt]
\eqntext{\mbox{if } N_{n(t+h)}-N_{nt}>1,}
\end{eqnarray}
and $\hat\mu_{\ri}=0$ otherwise,
and
%
\begin{eqnarray}
\hat\sigma^2_{\ri} =\hat\sigma^2_{\ri}(nt,nh)
=\frac{1}{N_{n(t+h)}-N_{nt}-2}\sum_{i=N_{nt}+2}^{N_{n(t+h)}} (
\xi_i-\hat\mu )^2
\nonumber\\[-8pt]\\[-8pt]
\eqntext{\mbox{if } N_{n(t+h)}-N_{nt}>2,}
\end{eqnarray}
and $\hat\sigma^2_{\ri}=0$ otherwise.

%
\begin{lemm}\label{lemmconvestimu}
Let $\Phi$ be an RPVV with associated mean $\mu$.
Let $T>0$, $h\in(0,T/2]$ and further $\hat\mu_{\ler}$ and $\hat\mu_{\ri}$
be defined as in (\ref{muschaetzer1}). Then it holds in $(D[h,T-h],
d_{\llVert  \cdot\rrVert  })$ a.s.~as $n\to\infty$ that
\begin{eqnarray*}
\bigl(\hat\mu_{\ler}(nt,nh) \bigr)_{t\in\tau_h} &\longrightarrow& (\mu
)_{t\in\tau_h},
\qquad
\bigl(\hat\mu_{\ri}(nt,nh)\bigr)_{t\in\tau_h}
\longrightarrow (\mu)_{t\in\tau_h}.
\end{eqnarray*}
\end{lemm}

\begin{pf}
Conditions (\ref{nullconda}) and (\ref{nullcondc}) imply
Kolmogorov's conditions (\ref{kol1})--(\ref{kol3}) with $\xi_i^2$ there
replaced by $\xi_i$. Hence, we have the SLLN for $\{\xi_i\}_{i\ge1}$.
Lemmas~\ref{lemm4}~and~\ref{lemm5} imply the strong consistency
for every fixed $t$, that is, almost surely as $n\to\infty$
%
\begin{equation}
\label{konsistenzmut} \hat{\mu}_{\ri}(nt,nh) =\frac{1}{N_{n(t+h)}-N_{nt}-1}\sum
_{i=N_{nt}+2}^{N_{n(t+h)}}\xi_i\longrightarrow\mu.
\end{equation}
Applying Slutsky's theorem with Lemma~\ref{lemm4}, we obtain for every
$t$ a.s.~as $n\to\infty$
%
\begin{equation}
\label{convmuaehnlich} \frac{\mu}{nh} \sum_{i=N_{nt}+2}^{N_{n(t+h)}}
\xi_i \longrightarrow\mu.
\end{equation}

In particular, the a.s. convergence holds for finitely many $t$ simultaneously.
In order to show the uniform a.s. convergence of $(\hat{\mu
}_{\ri}(nt,nh))$, we first show that in $(D[0,T-h],d_{\llVert  \cdot\rrVert  })$ it
holds a.s.~as $n\to\infty$ that
%
\begin{equation}
\label{uniformekonvergenzxi} \Biggl(\frac{\mu}{nh} \sum_{i=N_{nt}+2}^{N_{n(t+h)}}
\xi_i \Biggr)_{t\in
[0,T-h]} \longrightarrow(\mu)_{t\in[0,T-h]}.
\end{equation}
Note that as in the proof of Lemma~\ref{lemmconvnt}, for (\ref
{uniformekonvergenzxi}) it is sufficient to show that almost surely
%
\begin{eqnarray}\label{uniformekonvergenzxi2}
\lim_{n\to\infty} \sup_{t\in[0,T-h]}
\frac{\mu}{nh} \sum_{i=N_{nt}+2}^{N_{n(t+h)}}
\xi_i &\le& \mu \quad\mbox{and}
\nonumber\\[-9pt]\\[-9pt]\nonumber
\lim_{n\to\infty} \inf_{t\in[0,T-h]} \frac{\mu}{nh} \sum_{i=N_{nt}+2}^{N_{n(t+h)}}
\xi_i &\ge& \mu.
\end{eqnarray}
We use the same decomposition of the interval $(0,nT]$ into equidistant
sections of length $n \varepsilon$ as in the proof of Lemma~\ref
{lemmconvnt}. In order to see the left inequality of~(\ref{uniformekonvergenzxi2}), let $\varepsilon>0$. Since the life times
are nonnegative, with the notation (\ref{rn271113}), we can bound
\begin{eqnarray*}
&& \sup_{t\in[0,T-h]} \frac{\mu}{nh} \sum
_{i=N_{nt}+2}^{N_{n(t+h)}}\xi_i
\\[-2pt]
&&\qquad \le \max
_{k=0,1,\ldots,\lceil T/\varepsilon\rceil} \frac{\mu}{nh} \sum_{i=N_{kn\varepsilon}}^{N_{kn\varepsilon+n\llvert   \lceil h/\varepsilon\rceil
\rrvert  \varepsilon}}
\xi_i
\\
&&\qquad \le \max_{k=0,1,\ldots,\lceil T/\varepsilon\rceil} \frac{\mu}{nh} \sum
_{i=N_{kn\varepsilon+nh}}^{N_{kn\varepsilon+n\llvert   \lceil h/\varepsilon
\rceil\rrvert  \varepsilon}}\xi_i + \max
_{k=0,1,\ldots,\lceil T/\varepsilon\rceil} \frac{\mu}{nh} \sum_{i=N_{kn\varepsilon}}^{N_{kn\varepsilon+nh }}
\xi_i
\\
&&\qquad \le \frac{\mu}{h} \bigl(\bigl\llvert \lceil h/\varepsilon\rceil\bigr
\rrvert \varepsilon -h \bigr) + \max_{k=0,1,\ldots,\lceil T/\varepsilon\rceil} \frac{\mu}{nh}
\sum_{i=N_{kn\varepsilon}}^{N_{kn\varepsilon+nh }}\xi_i.
\end{eqnarray*}
The first summand in the previous line is independent of $n$ and tends
to zero as $\varepsilon\downarrow0$. Further, for every $\varepsilon
>0$, the second summand converges to $\mu$ a.s. as \mbox{$n \to\infty$},
according to equation (\ref{convmuaehnlich}).
Therefore, the first inequality in (\ref{uniformekonvergenzxi2})
holds. The second inequality in (\ref{uniformekonvergenzxi2}) can be
shown similarly, hence, (\ref{uniformekonvergenzxi}) holds.
In particular, we obtain the convergence in $(D[h,T-h],d_{\llVert  \cdot\rrVert  })$.

By Slutsky's theorem, (\ref{uniformekonvergenzxi}) and Lemma~\ref
{lemmconvnt} yield in $(D[h,T-h],d_{\llVert  \cdot\rrVert  })$ a.s.~as $n\to\infty
$ that
%
\begin{equation}
\label{uniformmu} \Biggl( \frac{1}{N_{n(t+h)}-N_{nt}-1}\sum_{i=N_{nt}+2}^{N_{n(t+h)}}
\xi_i \Biggr)_{t\in\tau_h} \longrightarrow(\mu)_{t\in\tau_h},
\end{equation}
which is the uniform a.s. consistency of $(\hat\mu_{\ri})_{t\in\tau_h}$.
In the same way we can conclude the uniform a.s. consistency of $(\hat\mu_{\ler})_{t\in\tau_h}$.
\end{pf}

Now we show the uniform a.s.~convergence of variance estimators.

%
\begin{lemm}\label{lemmconvestisigma}
Let $\Phi$ be an RPVV with variance $\sigma^2$.
Let $T>0$, $h\in(0,T/2]$ and further $\hat\sigma_{\ler}^2$ and $\hat\sigma_{\ri}^2$ be defined as in (\ref{sigmaschaetzer1}). Then in
$(D[h,T-h], d_{\llVert  \cdot\rrVert  })$ a.s.~as $n\to\infty$ we have
\[
\bigl(\hat\sigma_{\ler}^2(nt,nh) \bigr)_{t\in\tau_h}
\longrightarrow \bigl(\sigma^2\bigr)_{t\in\tau_h}, \qquad \bigl(\hat\sigma_{\ri}^2(nt,nh) \bigr)_{t\in\tau_h} \longrightarrow
\bigl(\sigma^2\bigr)_{t\in\tau_h}.
\]
\end{lemm}

\begin{pf}
For $N_{n(t+h)}-N_{nt}>2$ we decompose
\begin{eqnarray*}
\hat\sigma^2_{\ri}(nt,nh) & =& \frac{1}{N_{n(t+h)}-N_{nt}-2}\sum
_{i=N_{nt}+2}^{N_{n(t+h)}} (\xi_i-\hat\mu_{\ri} )^2
\\
& =& \frac{1}{N_{n(t+h)}-N_{nt}-2}\sum_{i=N_{nt}+2}^{N_{n(t+h)}} \xi
_i^2
\\
&&{} + \Biggl[ -2\hat\mu_{\ri} \Biggl(\frac{1}{N_{n(t+h)}-N_{nt}-2}\sum
_{i=N_{nt}+2}^{N_{n(t+h)}} \xi_i \Biggr) + \hat\mu_{\ri}^2 \Biggr].
\end{eqnarray*}
The expression in the squared brackets as a process in $t\in\tau_h$
converges to $(-\mu^2)_{t\in\tau_h}$ a.s.~in $(D[h,T-h],d_{\llVert  \cdot\rrVert
})$ due to the consistency of $(\hat\mu_{\ri})_{t\in\tau_h}$; see Lemma~\ref{lemmconvestimu}.

It remains to show that in $(D[h,T-h],d_{\llVert  \cdot\rrVert  })$ a.s.~as $n\to
\infty$
%
\begin{equation}
\label{uniformsumxiquad} \Biggl(\frac{1}{N_{n(t+h)}-N_{nt}-2}\sum_{i=N_{nt}+2}^{N_{n(t+h)}}
\xi _i^2 \Biggr)_{t\in\tau_h} \longrightarrow\bigl(
\sigma^2 + \mu^2\bigr)_{t\in\tau_h}.
\end{equation}
We abbreviate $\sigma_i^2:= \Var(\xi_i)$ and center
$(\xi_i^*)^2:=\xi_i^2 -(\sigma_i^2+\mu^2)$, so that
\begin{eqnarray*}
\frac{1}{N_{n(t+h)}-N_{nt}-2}\sum_{i=N_{nt}+2}^{N_{n(t+h)}}
\xi_i^2
&=& \frac{1}{N_{n(t+h)}-N_{nt}-2}\sum
_{i=N_{nt}+2}^{N_{n(t+h)}} \bigl(\xi_i^*
\bigr)^2
\\[-2pt]
&&{} + \Biggl[ \Biggl(\frac{1}{N_{n(t+h)}-N_{nt}-2}\sum_{i=N_{nt}+2}^{N_{n(t+h)}}
\sigma_i^2 \Biggr) +\mu^2 \Biggr].
\end{eqnarray*}
For fixed $t$ the term in the squared brackets converges to $\sigma^2 +
\mu^2$ a.s., as $n\to\infty$, by condition (\ref{nullcondd}) and
Lemma~\ref{lemm5}. Furthermore, condition (\ref{nullconde}) now
writes $(1/n) \sum_{i=1}^{n} (\xi_i^*)^2 \to0$ almost surely. Hence,
Lemma~\ref{lemm5} implies for fixed $t$
%
\begin{equation}
\label{limitxiquadrat} \frac{1}{N_{n(t+h)}-N_{nt}-2}\sum_{i=N_{nt}+2}^{N_{n(t+h)}}
\bigl(\xi_i^*\bigr)^2 \longrightarrow0\qquad\mbox{a.s.},
\end{equation}
as $n\to\infty$.

Thus, for finitely many $t$ we have the convergence in (\ref
{uniformsumxiquad}) a.s.~toward $\sigma^2 + \mu^2$.
In order to obtain the convergence in $(D[h,T-h],d_{\llVert  \cdot\rrVert  })$, we
proceed as in the proofs of Lemmas~\ref{lemmconvnt}~and~\ref{lemmconvestimu} and show a.s.~as $n\to\infty$ that
%
\begin{equation}
\label{uniformxiiquad3} \Biggl(\frac{\mu}{nh}\sum_{i=N_{nt}+2}^{N_{n(t+h)}}
\xi_i^2 \Biggr)_{t\in
\tau_h} \longrightarrow
\sigma^2+\mu^2.
\end{equation}
We again prove this claim even for $t\in[0,T-h]$.
Hence, it suffices to show a.s.~that
%
\begin{eqnarray}\label{uniformxiiquad2}
\lim_{n\to\infty} \sup_{t\in[0,T-h]}
\frac{\mu}{nh}\sum_{i=N_{nt}+2}^{N_{n(t+h)}}
\xi_i^2 &\le&\sigma^2+\mu^2 \quad\mbox {and}
\nonumber\\[-9pt]\\[-9pt]\nonumber
\lim_{n\to\infty} \inf_{t\in[0,T-h]}
\frac{\mu}{nh}\sum_{i=N_{nt}+2}^{N_{n(t+h)}}
\xi_i^2 &\ge&\sigma^2+\mu^2.
\end{eqnarray}
As in the previous proofs, for an $\varepsilon>0$, we decompose the
time interval $[0,nT]$ into equidistant sections of length $n\varepsilon
$ and, with notation (\ref{rn271113}), bound
\begin{eqnarray*}
&& \sup_{t\in[0,T-h]} \frac{\mu}{nh}\sum
_{i=N_{nt}+2}^{N_{n(t+h)}} \xi _i^2
\\
&&\qquad  \le
\max_{k=0,1,\ldots,\lceil T/\varepsilon\rceil} \frac{\mu
}{nh}\sum_{i=N_{kn\varepsilon}}^{N_{kn\varepsilon+n\llvert   \lceil
h/\varepsilon\rceil\rrvert  \varepsilon}}
\xi_i^2
\\
&&\qquad  \le\max_{k=0,1,\ldots,\lceil T/\varepsilon\rceil} \frac{\mu
}{nh}\sum
_{i=N_{kn\varepsilon+nh}}^{N_{kn\varepsilon+n\llvert   \lceil
h/\varepsilon\rceil\rrvert  \varepsilon}} \xi_i^2
+ \max
_{k=0,1,\ldots,\lceil T/\varepsilon\rceil} \frac{\mu}{nh}\sum_{i=N_{kn\varepsilon}}^{N_{kn\varepsilon+nh }}
\xi_i^2.
\end{eqnarray*}
For $\delta:=\lceil h/\varepsilon\rceil\varepsilon-h+\varepsilon$ we
find a.s.~for $n\to\infty$,
\[
\max_{k=0,1,\ldots,\lceil T/\varepsilon\rceil} (N_{kn\varepsilon+ n\llvert
\lceil h/\varepsilon\rceil\rrvert  \varepsilon} - N_{kn\varepsilon
+nh})/(\delta n/
\mu) \to1.
\]
Then for $n\to\infty$ the first summand in the latter display converges
to $(\delta/h)(\sigma^2+\mu^2)$ a.s. and the second summand to $\sigma
^2+\mu^2$ a.s., since we have the convergence (\ref{uniformsumxiquad}) for finitely many $t$.
Since $\delta$ can be chosen arbitrarily small, we find the first
inequality of (\ref{uniformxiiquad2}). The second follows
analogously, and the convergence in (\ref{uniformxiiquad3})
follows. There, we exchange the normalization according to Lemma~\ref
{lemmconvnt} and obtain (\ref{uniformsumxiquad}). Thus, the
a.s.~uniform consistency of the variance estimator $(\hat\sigma
_{\ri}^2)_{t\in\tau_h}$ is proven.
The uniform a.s. convergence of $(\hat\sigma_{\ler}^2)_{t\in\tau_h}$ is
obtained analogously.
\end{pf}

\subsection{Proof of Theorem \texorpdfstring{\protect\ref{hauptaussagegeneralisiert}}{3.1}}\label{proofhauptaussagegeneralisiert}
Finally, we put the pieces of the previous subsections together to
prove Theorem~\ref{hauptaussagegeneralisiert}:

\begin{pf*}{Proof of Theorem~\ref{hauptaussagegeneralisiert}}
Let $\Phi$ be an RPVV with associated parameters $\mu$ and $\sigma^2$
and conditions as is Theorem~\ref{hauptaussagegeneralisiert}. The
associated counting process is denoted by $(N_t)_{t\ge0}$; cf.~(\ref
{defnt}). Further, let $T>0$ and $h\in(0,T/2]$ denote a window size.

From Proposition~\ref{proprescaledcount} we have that the
normalization of $(N_{t})_{t\ge0}$ given by
\[
Z_t^{(n)} = \frac{N_{n t} - \nicefrac{n t}{\mu}}{\sqrt{n}\sqrt{\nicefrac
{\sigma^2}{\mu^3}}}, \qquad t\ge0,
\]
converges, as $n\to\infty$ in distribution in $(D[0,\infty), d_{\mathrm{SK}})$
to a standard Brownian motion:
%
\begin{equation}
\label{convresccountbeweis} \bigl(Z_t^{(n)}\bigr)_{t\ge0}
\stackrel{d} {\longrightarrow} (W_t)_{t \ge0}.
\end{equation}

Now, for technical reasons we define an auxiliary process, for $t\ge0$
and $h\in(0,T/2]$, by
\[
\Gamma_{t,h}^{(n)}:= \frac{
(N_{n(t+h)}-N_{nt})-(N_{nt}-N_{n(t-h)})}{\sqrt{2hn\sigma^2/\mu^3}}.
\]
In comparison with the $G_{t,h}^{(n)}$ defined in (\ref
{ghtextension}), note that the $\Gamma_{t,h}^{(n)}$ are normalized
deterministically with the order of the estimator $\hat{s}$ used for
normalization in (\ref{ghtextension}). Now, we apply the continuous
mapping theorem as follows:
the map $\varphi\dvtx  (D[0,\infty),d_{\mathrm{SK}}) \to(D[h,T-h],d_{\mathrm{SK}})$ defined by
\[
f(t)\stackrel{\varphi} {\longmapsto} \frac{(f(t+h)-f(t)) -
(f(t)-f(t-h))}{\sqrt{2h}}
\]
is continuous.
With the process $(Z^{(n)})_{t\ge0}$ defined in (\ref{rn261113}), we
have\break $\varphi((Z^{(n)})_{t\ge0})=(\Gamma_{h,t}^{(n)})_{t\in\tau_h}$.
Furthermore, the process $(L_{h,t})_{t\in\tau_h}$ defined in (\ref
{limitprocess}) is distributed as $\varphi((W_t)_{t\ge0})$ with a
standard Brownian motion $(W_t)_{t\ge0}$. Hence, the convergence~(\ref
{convresccountbeweis}) and the continuous mapping theorem imply the
weak convergence in Skorokhod topology of $(\Gamma_{h,t}^{(n)})_{t\in
\tau_h}$ to $(L_{h,t})_{t\in\tau_h}$.\vspace*{1pt}

By Proposition~\ref{propconvergenceshat} we have in $(D[h,T-h],d_{\llVert  \cdot\rrVert  })$ a.s.~as $n\to\infty$ that
%
\begin{equation}
\label{rn261113b} \biggl(\frac{\hat s(nt,nh)^2}{n} \biggr)_{t\in\tau_h}\longrightarrow
\biggl(\frac{2h\sigma^2}{\mu^3} \biggr)_{t\in\tau_h}.
\end{equation}
Since we have the relation
\[
G^{(n)}_{h,t}= \frac{\sqrt{2nh\sigma^2/\mu^3}}{\hat{s}(nt,nh)}\Gamma ^{(n)}_{h,t},
\]
we conclude by Slutsky's theorem with (\ref{rn261113b}) that in
$(D[h,T-h],d_{\mathrm{SK}})$ it holds
\[
\bigl(G^{(n)}_{h,t}\bigr)_{t \in\tau_h} \stackrel{d} {
\longrightarrow} (L_{h,t})_{t\in\tau_h}.
\]
This is the assertion.
\end{pf*}
\end{appendix}

\section*{Acknowledgments}
We thank Brooks Ferebee and Markus Bingmer for stimulating discussions
and helpful ideas. We are grateful to Rudolf Gr\"ubel and G\"otz
Kersting for technical advice.

\begin{supplement}
\stitle{Supplement to ``A multiple filter test for the detection of rate changes in renewal processes with varying variance''}
\slink[doi]{10.1214/14-AOAS782SUPP} 
\sdatatype{.r}
\sfilename{aoas782\_supp.r}
\sdescription{We provide the R-Code for the multiple filter algorithm.}
\end{supplement}

%

\printaddresses
\end{document}